\documentclass[prx,reprint,superscriptaddress]{revtex4-2}
\pdfoutput=1
\usepackage{braket,dsfont,subfigure,amsfonts,amssymb,bm,graphicx,float,amsmath,hyperref,natbib}
\usepackage{soul}
\usepackage[utf8]{inputenc}
\usepackage[english]{babel}
\usepackage[T1]{fontenc}
\usepackage{tikz}
\usepackage{lipsum}

\begin{document}
\newcommand{\rrangle}{\rangle\!\rangle}
\newcommand{\llangle}{\langle\!\langle}
\newcommand{\rrpipe}{|}
\newcommand{\llpipe}{|}
\newcommand{\llvert}{\left|\hspace{-0.5mm}\left|}
\newcommand{\rrvert}{\right|\hspace{-0.5mm}\right|}
\newcommand{\sket}[1]{\ensuremath{\llpipe#1\rrangle}}
\soulregister\cite7
\soulregister\eqref7
\soulregister\ref7
\title{Quantum error mitigation via matrix product operators}

\author{Yuchen Guo}
\affiliation{State Key Laboratory of Low Dimensional Quantum Physics and Department of Physics, Tsinghua University, Beijing 100084, China}
\author{Shuo Yang}
\email{shuoyang@tsinghua.edu.cn}
\affiliation{State Key Laboratory of Low Dimensional Quantum Physics and Department of Physics, Tsinghua University, Beijing 100084, China}
\affiliation{Frontier Science Center for Quantum Information, Beijing 100084, China}
\affiliation{Hefei National Laboratory, Hefei 230088, China}

\begin{abstract}
In the era of noisy intermediate-scale quantum (NISQ) devices, the number of controllable hardware qubits is insufficient to implement quantum error correction (QEC).
As an alternative, quantum error mitigation (QEM) can suppress errors in measurement results via repeated experiments and postprocessing of data.
Typical techniques for error mitigation, e.g., the quasi-probability decomposition method, ignore correlated errors between different gates.
Here, we introduce a QEM method based on the matrix product operator (MPO) representation of a quantum circuit that can characterize the noise channel with polynomial complexity.
Our technique is demonstrated on a $\rm{depth}=20$ fully parallel quantum circuit of up to $N_q=20$ qubits undergoing local and global noise.
The circuit error is reduced by several times with only a small bond dimension $D^{\prime} = 1$ for the noise channel.
The MPO representation increases the accuracy of modeling noise without consuming more experimental resources, which improves the QEM performance and broadens its scope of application.
Our method is hopeful of being applied to circuits in higher dimensions with more qubits and deeper depth.
\end{abstract}
\maketitle

\section{Introduction}
The idea of quantum supremacy \cite{Preskill2012, Arute2019} is to take advantage of the exponential complexity of quantum systems to build information processing devices that exceed the power of classical supercomputers.
However, universal fault-tolerant quantum computation \cite{Preskill1997}, which requires the manipulation of millions or more qubits to implement quantum error correction (QEC) \cite{Shor1995, Calderbank1996}, is beyond our reach for the time being.
State-of-the-art hardware composed of noisy intermediate-scale quantum (NISQ) devices typically contains hundreds of qubits with error rates on the order of $10^{-3}$ \cite{Endo2021}.
Many interesting quantum-classical hybrid algorithms can be run on these devices, including variational quantum eigensolver (VQE) \cite{Peruzzo2014, McClean2016, Kandala2017}, variational quantum simulation (VQS) \cite{Li2017, Yuan2019}, etc.
To prevent error accumulation in NISQ devices, many approaches for quantum error mitigation (QEM) are proposed to suppress errors in measurement results via data postprocessing.

Previously studied QEM methods include error extrapolation \cite{Li2017,Temme2017,Otten2019}, quasi-probability method \cite{Temme2017,Endo2018}, quantum subspace expansion \cite{McClean2017}, symmetry verification \cite{Bonet2018,McArdle2019}, and several learning-based approaches \cite{Strikis2021,Czarnik2021,zhang2021}.
Different techniques can be combined, e.g., combinations of error extrapolation, quasi-probability, and symmetry verification \cite{Cai2021}.
Experimental QEMs are reported in a trapped-ion system \cite{Zhang2020} and a superconducting system \cite{Kandala2019}.

State-of-the-art QEM techniques, such as the quasi-probability method, try to mitigate noise for every gate independently.
It leads to ignorance of temporally or spatially correlated errors, limiting the performance of QEM due to the inaccurate characterization of noise models \cite{Cao2021}.
A scalable QEM method capable of dealing with models beyond localized and Markovian noise remains to be found.

In this Letter, we propose a QEM framework based on the matrix product operator (MPO) representation of a noisy quantum circuit and its noise channel.
We use the quantum process tomography (QPT) technique proposed by Torlai \textit{et al.} \cite{Torlai2020} to obtain the MPO representation and introduce a variational algorithm to calculate the inverse of the noise channel.
Our QEM approach can characterize the noise model with only polynomial complexity, which facilitates the design of correcting circuits capable of mitigating noise.

\begin{figure*}
    \centering
    \includegraphics[width = 0.95\linewidth]{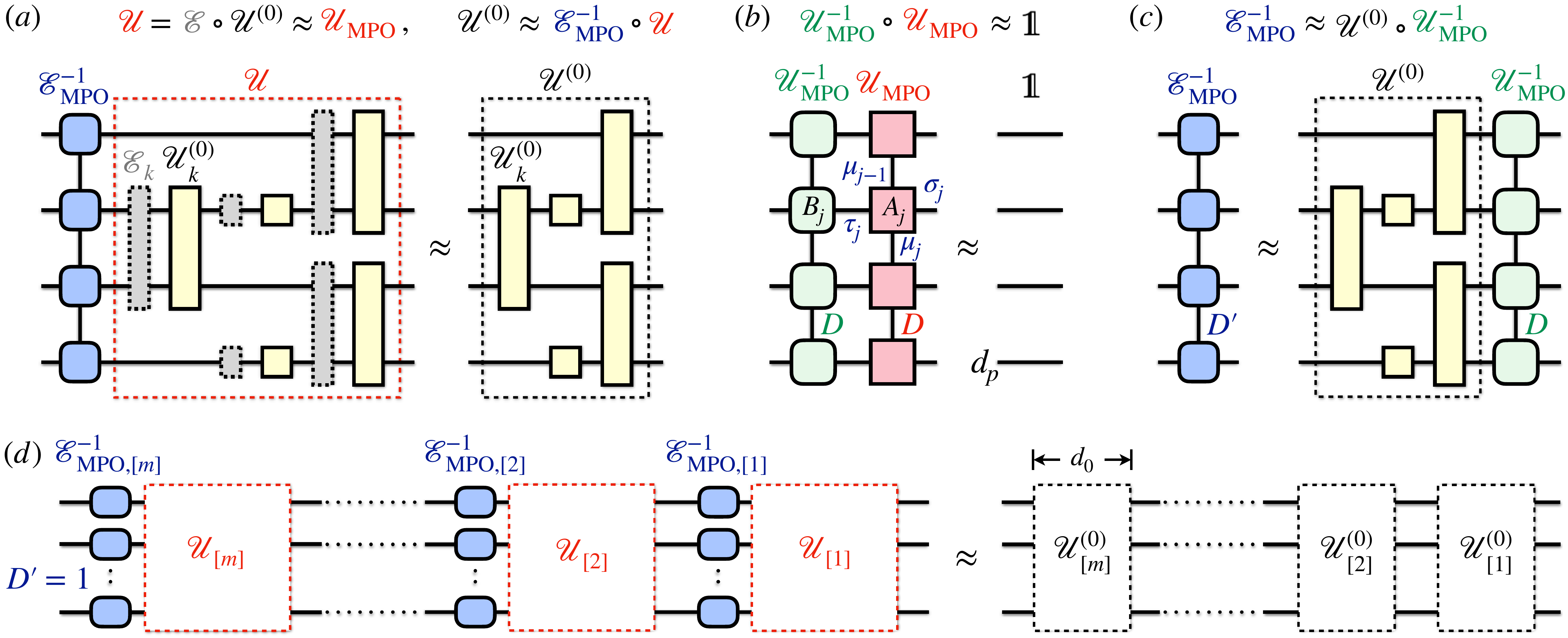}
    \caption{(Color online) (a) The schematic diagram of our QEM method based on MPO.
    We first use an MPO to represent the noisy quantum circuit $\mathcal{U}_{\rm{MPO}}$.
    Then we calculate the inverse noise channel $\mathcal{E}^{-1}_{\rm{MPO}}$, which is applied after $\mathcal{U}$ to compensate for the error and to restore the ideal circuit $\mathcal{U}^{(0)}$.
    (b) Our variational MPO-inverse method.
    We calculate the inverse of an MPO-represented quantum channel $\mathcal{U}_{\rm{MPO}}$, which is parameterized as an MPO $\mathcal{U}^{-1}_{\rm{MPO}}$ with the same bond dimension $D$.
    (c) Calculation of the inverse noise channel $\mathcal{E}^{-1}_{\rm{MPO}}$ via MPO contraction and truncation methods, whose bond dimension is $D^{\prime}$.
    (d) A deep circuit is divided into $m$ parts, each with $d_0$ layers. One may apply our QEM method on each part, where $\mathcal{E}^{-1}_{\textrm{MPO}, [k]}$ is truncated to $D^{\prime}=1$ and simulated by single-qubit gates.}
    \label{Fig: MPO_QEM}
\end{figure*}

\section{Quasi-probability method}
The quasi-probability method was first introduced by Temme \textit{et al.} \cite{Temme2017} for specific noise models, and was then generalized by Endo \textit{et al.} \cite{Endo2018} to any localized and Markovian errors with the help of quantum gate set tomography (GST) \cite{Merkel2013,greenbaum2015,Nielsen2021}.
We use $\mathcal{U}^{(0)}_k$ to denote the quantum channel of the $k$-th ideal gate in the circuit, while the actual noisy gate $\mathcal{U}_k$ is denoted as $\mathcal{U}_k = \mathcal{E}_k\circ \mathcal{U}^{\left(0\right)}_k$ with $\mathcal{E}_k$ specifying the noise channel.
One can implement GST to characterize the noise channel $\mathcal{E}_k$, then apply $\mathcal{E}^{-1}_k$ after each gate to invert the noise effect
\begin{align}
    \mathcal{E}^{-1}_k\circ\mathcal{U}_k=\mathcal{E}^{-1}_k\circ \mathcal{E}_k\circ \mathcal{U}^{(0)}_k = \mathcal{U}^{(0)}_k.
\end{align}

In practice, one may use Monte Carlo sampling to simulate the operator $\mathcal{E}^{-1}_k$.
With a universal set of real gate channels $\mathcal{B}_{i_k}$, which we assume to be complete and can be realized in experiments, one decomposes the inverse noise channel as $\mathcal{E}^{-1}_k = \sum_{i_k}{q_{i_k}\mathcal{B}_{i_k}}$.
Consequently, by randomly applying $\mathcal{B}_{i_k}$ after $\mathcal{U}_k$ with probability $p_{i_k} = \left|q_{i_k}\right|/C_k$, one can obtain the ideal measurement result for any observable $\mathcal{O}$
\begin{align}
    \begin{aligned}
        \braket{\mathcal{O}}^{(0)} &= \textrm{Tr}{\left[\mathcal{O} \mathcal{U}^{(0)}_k\left(\rho\right)\right]} = \textrm{Tr}{\left[\mathcal{O} \mathcal{E}_k^{-1}\circ\mathcal{U}_k\left(\rho\right)\right]}\\
        &= C_k\sum_{i_k}{\textrm{sgn}\left(q_{i_k}\right)p_{i_k}\textrm{Tr}{\left[\mathcal{O}\mathcal{B}_{i_k} \circ \mathcal{U}_k\left(\rho\right)\right]}},
    \end{aligned}
\end{align}
where $C_k = \sum_{i_k}{\left|q_{i_k}\right|}$ is the normalization factor.
We note that $C_k^2$ labels the sampling cost to simulate $\mathcal{E}^{-1}_k$ \cite{Takagi2021}, which is related to its physical implementability \cite{Jiang2021}.
Typically $C_k \approx 1+b\varepsilon_k$ with a positive number $b\lesssim2$ \cite{Endo2018}, where $\varepsilon_k$ is the error rate of the $k$-th gate.

For the entire quantum circuit $\prod_{k=1}^{N_g}{\mathcal{U}^{\left(0\right)}_{k}}$, the ideal process is represented as
\begin{align}
    \mathcal{U}^{(0)} = \prod_{k=1}^{N_g}{\mathcal{U}^{\left(0\right)}_{k}} = C_{\rm{tot}}\sum_{\vec{i}}\textrm{sgn}\left(q_{\vec{i}}\right)p_{\vec{i}}\prod_{k=1}^{N_g}{\mathcal{B}_{i_k}}\circ \mathcal{U}_k,
\end{align}
with $\vec i=(i_1,i_2,\dots,i_{N_g})$, $q_{\vec i}=\prod_{k=1}^{N_g}q_{i_k}$, and $p_{\vec i}=\prod_{k=1}^{N_g}p_{i_k}$.
The entire variance amplification becomes $C_{\rm{tot}}^2=\prod_{k=1}^{N_g}C_k^2$, which calls for $C^2_{\rm{tot}}$ times more samples to achieve the same precision.

The application of the above method requires all the noise channels in a quantum circuit to be local and Markovian, omitting the correlation between errors of different gates.
In other words, it fails to treat correlated noise, such as the crosstalk noise between two adjacent gates or the global depolarizing noise across the entire system.

\section{Quantum error mitigation via matrix product operators}
The imperfect characterization of the noise channel in standard QEM techniques motivates us to treat the noise model differently, e.g., by dealing with several layers of a quantum circuit as a whole to take these global and non-Markovian errors into account.
With tensor network (TN) methods \cite{Verstraete2008,Orus2014,Bridgeman2017,Cirac2021}, such a task can be accomplished efficiently.
In particular, TN provides an intuitive comprehension and a simple representation of the intrinsic entanglement structure for many-body wavefunctions.
The total number of variational parameters and the computational cost is polynomial with system size $N_q$.

TN family has been applied to some data-driven reconstruction tasks in quantum computation before, e.g., quantum state tomography (QST) via matrix product states (MPS) \cite{Cramer2010,Baumgratz2013,Baumgratz_2013,Lanyon2017,Wang2020} and quantum process tomography (QPT) via matrix product operators (MPO) \cite{Guo2020,Torlai2020}.
These TN-based methods require only polynomially increasing resources for experiments and computation, while standard procedures for QST and QPT scale exponentially with system size.
Inspired by these studies and standard TN algorithms, we propose to perform QEM via matrix product operators.

The schematic diagram of our MPO-based error mitigation approach is shown in Fig. \ref{Fig: MPO_QEM}.
We consider an ideal quantum circuit $\mathcal{U}^{\left(0\right)}$, whose corresponding real circuit behaves as $\mathcal{U} = \mathcal{E}\circ \mathcal{U}^{\left(0\right)}$ with all errors in the circuit characterized by a noise channel $\mathcal{E}$.
We assume that $\mathcal{U}$ has a shallow depth and thus has an efficient MPO representation \cite{Wood2015, Bridgeman2017}.
We first apply QPT on the noisy quantum circuit to obtain its MPO representation $\mathcal{U}_{\textrm{MPO}}$.
Then we calculate the inverse of the noise channel $\mathcal{E}^{-1}_{\rm{MPO}} = \mathcal{U}^{\left(0\right)}\circ \mathcal{U}^{-1}_{\textrm{MPO}}$ via the MPO-inverse method to be introduced later, as shown in Fig. \ref{Fig: MPO_QEM}(b)(c).
Finally, one may design corresponding quantum circuits to implement the linear map $\mathcal{V} \approx \mathcal{E}^{-1}_{\rm{MPO}}$ after $\mathcal{U}$ to null out the error, i.e.,
\begin{align}
    \mathcal{V}\circ\mathcal{U}\approx\mathcal{E}^{-1}_{\textrm{MPO}}\circ\mathcal{E}\circ\mathcal{U}^{(0)}\approx\mathcal{U}^{\left(0\right)}.
\end{align}

A necessary condition for implementing such a type of QEM approach is that the correcting circuit $\mathcal{V}$ itself will not induce more errors than the errors to be mitigated in the original circuit \cite{Cao2021}.
Therefore, we choose single-qubit gates to simulate $\mathcal{E}^{-1}$, whose typical error rate is about one order of magnitude smaller than that of two-qubit gates in state-of-the-art quantum devices.
In other words, we truncate the bond dimension of $\mathcal{E}^{-1}_{{\rm MPO}}$ to $D^{\prime} = 1$ and apply the corresponding single-qubit operations after the end of the circuit.

In the NISQ era, we may encounter deeper noisy quantum circuits showing quantum advantage, where the classical simulation for $\mathcal{E}^{-1}$ is intractable.
In this case, one can divide the deep circuit into several parts, and implement our proposed method on each part, as shown in Fig. \ref{Fig: MPO_QEM}(d).
Assume that the entire circuit is composed of $m$ parts, each with $d_0$ layers, which can be written as
\begin{align}
    \mathcal{U} = \mathcal{E}_{[m]}\circ\mathcal{U}^{(0)}_{[m]}\circ\cdots\circ\mathcal{E}_{[2]}\circ\mathcal{U}^{(0)}_{[2]}\circ\mathcal{E}_{[1]}\circ\mathcal{U}^{(0)}_{[1]}.
\end{align}
After implementing our QEM method, it will behave as
\begin{align}
    \prod_{k = 1}^{m} \mathcal{V}_{[k]}\circ \mathcal{U}_{[k]} \approx \prod_{k = 1}^{m} \mathcal{E}^{-1}_{{\rm MPO}, [k]}\circ\mathcal{U}_{[k]} \approx \prod_{k = 1}^{m} \mathcal{U}^{(0)}_{[k]} = \mathcal{U}^{(0)},
\end{align}
i.e., noise effects in the original quantum circuit are approximately canceled out.

The sampling cost to simulate a noise inverse $\mathcal{E}^{-1}$ is defined by its physical implementability.
In this regard, our method is upper bound by the standard quasi-probability method since we treat several sequential noise channels as a whole, which might cancel out mutually to some extent, as proved in Theorem 6,7 in \cite{Jiang2021}.

\section{MPO representation of noisy quantum circuits}
A general quantum channel has many equivalent mathematical representations.
We adopt the superoperator description in our QEM method \cite{Nielsen2021}.
An $N_q$-qubit quantum state $\rho$ in the form of a Hermitian $2^{N_q}\times 2^{N_q}$ matrix can be rearranged as a $4^{N_q}$-dimensional vector $\sket{\rho}$.
A quantum circuit $\mathcal{U}$ acts linearly on quantum states and is a completely-positive trace-preserving (CPTP) map $\sket{\rho} \mapsto \mathcal{U} \sket{\rho}$ \cite{Nielsen2009}.
One can use a $4^{N_q}\times 4^{N_q}$ matrix to represent a $N_q$-qubit quantum circuit in the superoperator form.

\begin{figure}
    \centering
    \includegraphics[width = 0.7\linewidth]{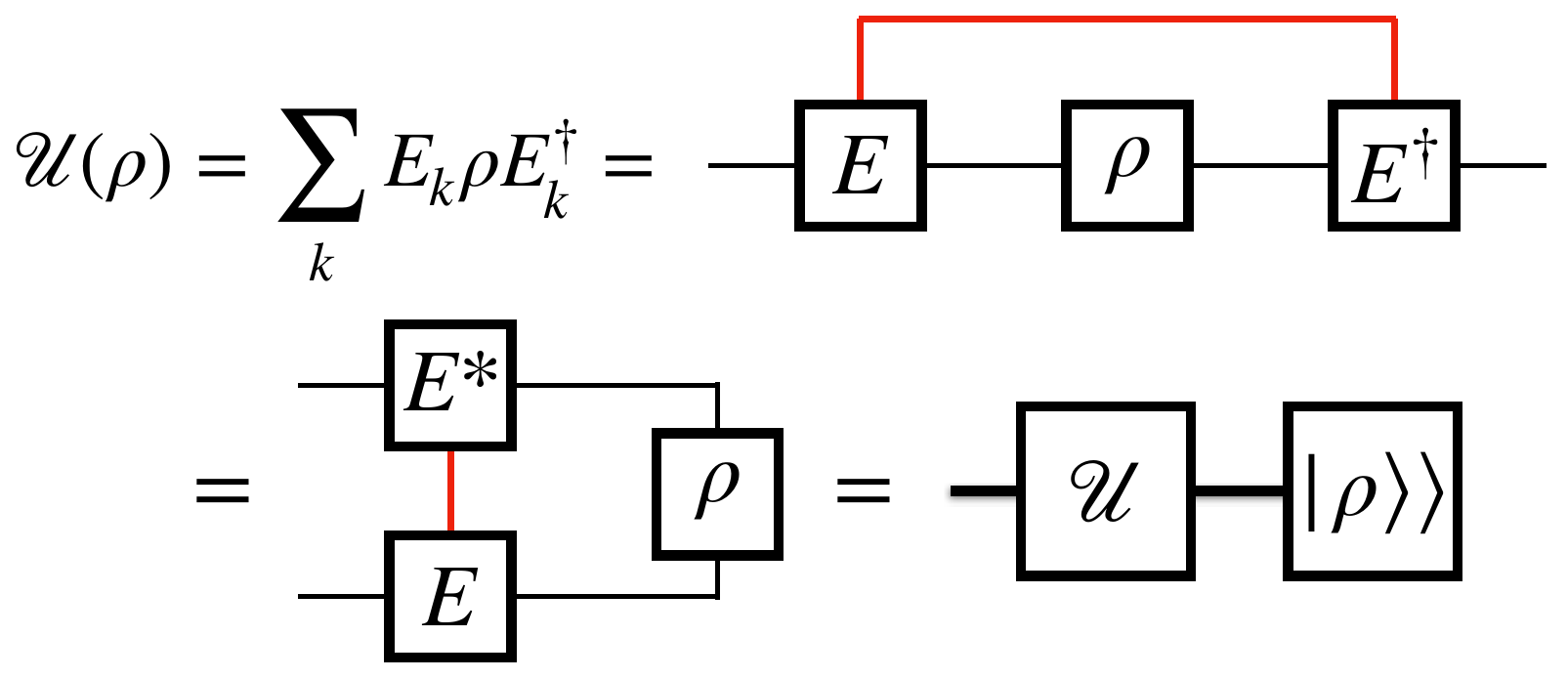}
    \caption{(Color online) The tensor representation of a quantum channel in its operator-sum form $\sum_k{E_k\rho E_k^{\dagger}}$.
    The contraction of the index $k$ (red line) corresponds to the summation of $E_k$.
    We group the two indices of $\rho$ to form a vector $\sket{\rho}$, then we obtain the superoperator form of $\mathcal{U}$.}
    \label{Fig: Kraus}
\end{figure}

To construct the superoperator for a quantum channel, we start with the so-called operator-sum representation \cite{Nielsen2009}, where a quantum channel $\mathcal{U}$ is represented as
\begin{align}
    \mathcal{U}\left(\rho\right) = \sum_k{E_k\rho E_k^{\dagger}}.
\end{align}
Here the operators $E_k$ are operation elements for $\mathcal{U}$ and satisfy the completeness relation $\sum_k{E_k^{\dagger}E_k} = 1$ which guarantees that $\mathcal{U}$ preserves the trace.
One can use the contraction of tensors to replace the summation in Fig. \ref{Fig: Kraus} \cite{Wood2015}.
Therefore, an $N_q$-qubit quantum circuit $\mathcal{U}$ can be represented by a $4^{N_q}\times 4^{N_q}$ matrix, which is just the superoperator form used in our method.

For example, a $Z$ gate in its superoperator form is
\begin{align}
    \mathcal{U}_Z = Z \otimes Z^* = \begin{bmatrix}
        1 & 0 & 0 & 0 \\
        0 & -1 & 0 & 0 \\
        0 & 0 & -1 & 0 \\
        0 & 0 & 0 & 1 \\
        \end{bmatrix}\:,
    \label{equ: super}
\end{align}
while the superoperator for a general quantum channel is
\begin{align}
    \mathcal{U} = \sum_k{E_k\otimes E_k^*}.
\end{align}

\begin{figure}
    \centering
    \includegraphics[width = 0.7\linewidth]{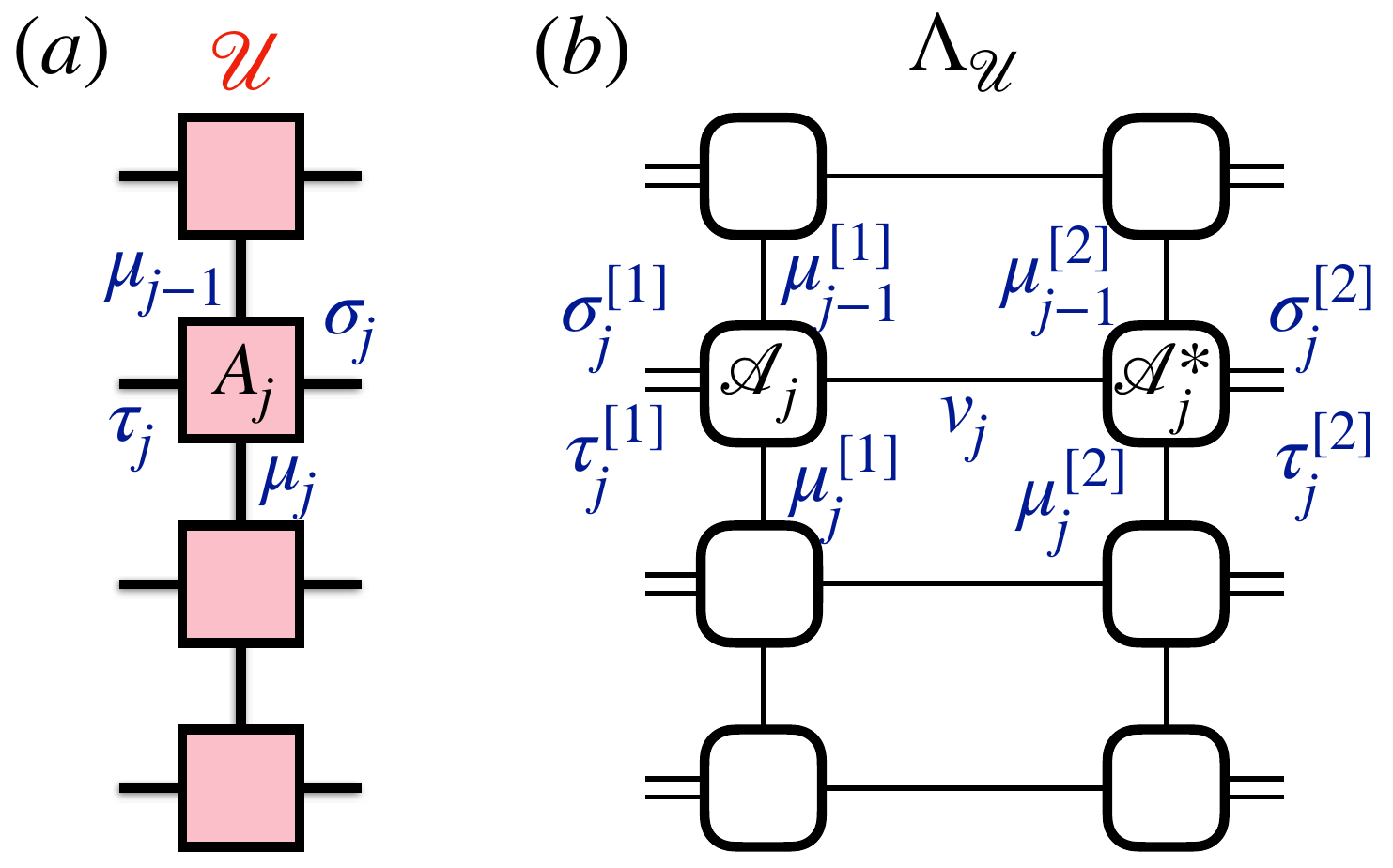}
    \caption{(Color online) Tensor network representations of a quantum channel $\mathcal{U}$. 
    (a) The superoperator $\mathcal{U}$ in its MPO form.
    (b) The Choi matrix $\Lambda_{\mathcal{U}}$ parameterized by a LPDO.}
    \label{Fig: LPDO}
\end{figure}

After separating the degrees of freedom at each site, a superoperator can further be considered as an MPO (or more precisely, a matrix product superoperator) with physical dimension $d_p=4$, i.e.,
\begin{align}
    \mathcal{U}^{\bm{\tau}}_{\bm{\sigma}}=\sum_{\{\bm{\mu}\}}\:\prod_{j=1}^N\:[A_j]^{\tau_j,\sigma_j}_{\mu_{j-1},\mu_{j}},
\end{align}
where $\bm{\sigma}=\{\sigma_j\}$ and $\bm{\tau}=\{\tau_j\}$ are respectively the input and output indices on each site, as shown in Fig. \ref{Fig: LPDO} (a).

Another mathematical form to describe a quantum channel is the Choi matrix $\Lambda$ \cite{Choi1975}, which also has a tensor network representation called locally-purified density operator (LPDO) \cite{Werner2016, Torlai2020}
\begin{align}
    \left[\Lambda_{\mathcal{U}}\right]^{\bm{\tau}^{[1]}, \bm{\tau}^{[2]}}_{\bm{\sigma}^{[1]}, \bm{\sigma}^{[2]}}=\sum_{\{\bm{\mu}^{[1]}, \bm{\mu}^{[2]}\}}\sum_{\{\bm{\nu}\}}\prod_{j=1}^N[\mathcal{A}_j]^{\tau_j^{[1]},\sigma_j^{[1]}}_{\mu_{j-1}^{[1]}, \nu_{j}, \mu_{j}^{[1]}}[\mathcal{A}_j^*]^{\tau_j^{[2]},\sigma_j^{[2]}}_{\mu_{j-1}^{[2]}, \nu_{j}, \mu_{j}^{[2]}}\:,
\end{align}
as shown in Fig. \ref{Fig: LPDO} (b). Here $\bm{\sigma}^{[i]}$ and $\bm{\tau}^{[i]}$ ($i = 1, 2$) are respectively the input and output indices, and $\bm{\sigma}^{[1]}, \bm{\sigma}^{[2]}$ are applied on each side of density matrices.

The conversion between the superoperator and the Choi matrix can be realized by just reshuffling the physical indices.
In the tensor network representation, to convert a Choi matrix parameterized by an LPDO into a superoperator in its MPO form, one needs to contract two local tensors on the same site and group corresponding physical and bond indices, i.e., 
\begin{align}
    [A_j]^{\tau_j,\sigma_j}_{\mu_{j-1},\mu_{j}} = \sum_{\nu_j}[\mathcal{A}_j]^{\tau_j^{[1]},\sigma_j^{[1]}}_{\mu_{j-1}^{[1]}, \nu_{j}, \mu_{j}^{[1]}}[\mathcal{A}_j^*]^{\tau_j^{[2]},\sigma_j^{[2]}}_{\mu_{j-1}^{[2]}, \nu_{j}, \mu_{j}^{[2]}}
\end{align}
with $\tau_j = \left\{\tau_j^{[1]}, \tau_j^{[2]}\right\}$, $\sigma_j = \left\{\sigma_j^{[1]}, \sigma_j^{[2]}\right\}$ and $\mu_j = \left\{\mu_j^{[1]}, \mu_j^{[2]}\right\}$.

\section{Inverse of matrix product operators}\label{Sec: MPO_inverse}
We now discuss how to calculate the inverse of a quantum channel $\mathcal{U}$ in its MPO representation, as shown in Fig. \ref{Fig: MPO_QEM}(b).
It is realized by minimizing the error
\begin{align}
    \begin{aligned}
        e &= \llvert\mathcal{U}^{\prime}\mathcal{U}-\mathds{1}\rrvert_{F}^2 = \textrm{Tr}\left[\left(\mathcal{U}^{\prime}\mathcal{U}-\mathds{1}\right)\left(\mathcal{U}^{\prime}\mathcal{U}-\mathds{1}\right)^{\dagger}\right]\\
        &= \textrm{Tr}{\left[\mathcal{U}^{\prime}\mathcal{U}\mathcal{U}^{\dagger}\mathcal{U}^{\prime\dagger}\right]} - \textrm{Tr}{\left[\mathcal{U}^{\dagger}\mathcal{U}^{\prime\dagger}\right]} - \textrm{Tr}{\left[\mathcal{U}^{\prime}\mathcal{U}\right]}+\textrm{Tr}{\left[\mathds{1}\right]}, \label{Equ: Mpo_inverse}
    \end{aligned}
\end{align}
where we update parameters of $\mathcal{U}^{\prime}$ to approach $\mathcal{U}^{-1}$, and $\llvert\dots\rrvert_F$ is the Frobenius norm of a matrix.

\begin{figure}
    \centering
    \includegraphics[width=\linewidth]{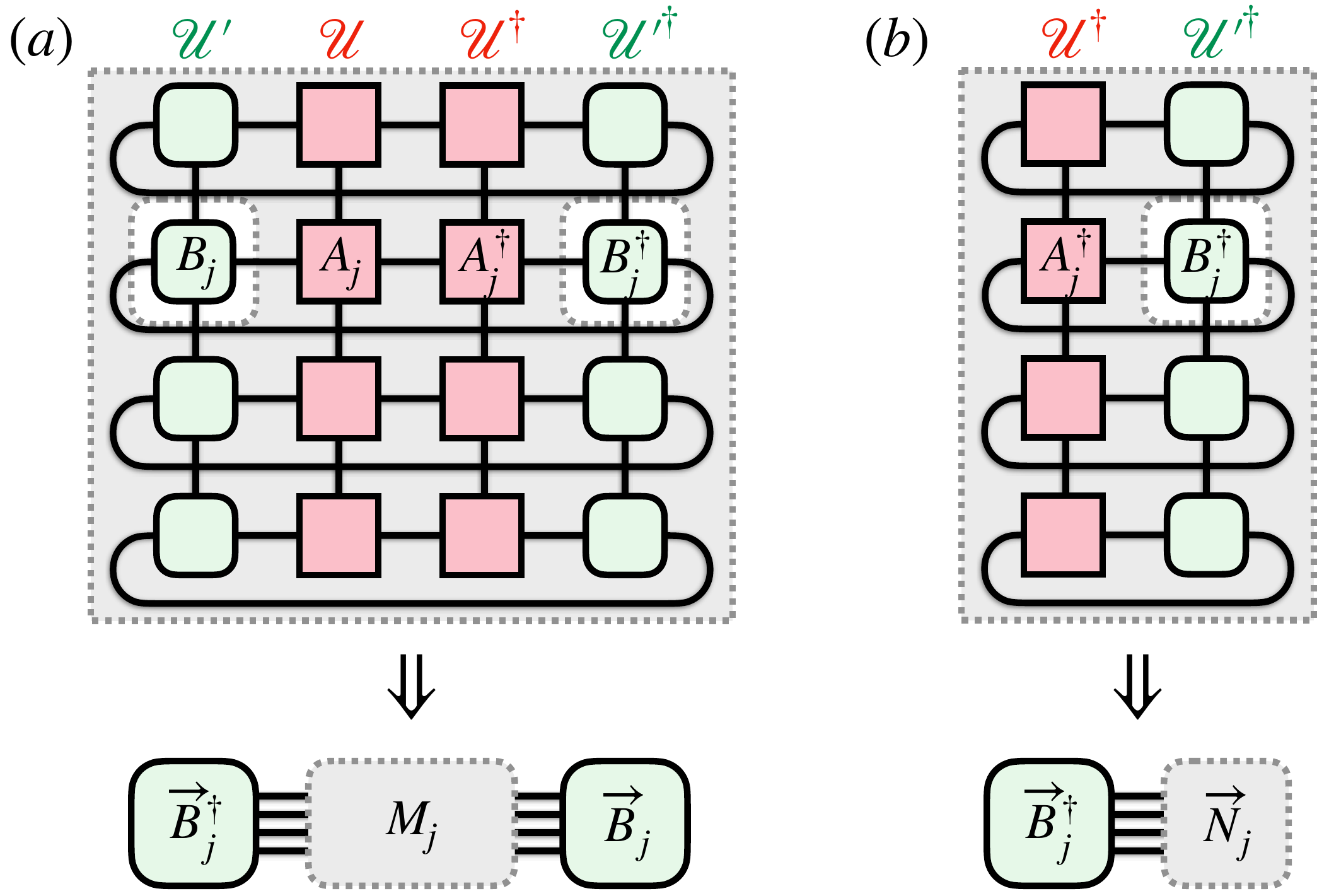}
    \caption{(Color online) The schematic diagram for the minimization of Eq. \eqref{Equ: Mpo_inverse}.
    (a) $M_j$ is the environment of $\vec{B}_j$ and $\vec{B}_j^{\dagger}$ in $\textrm{Tr}{\left[\mathcal{U}^{\prime}\mathcal{U}\mathcal{U}^{\dagger}\mathcal{U}^{\prime\dagger}\right]}$.
    (b) $\vec{N}_j$ is the environment of $\vec{B}_j^{\dagger}$ in $\textrm{Tr}{\left[\mathcal{U}^{\dagger}\mathcal{U}^{\prime\dagger}\right]}$.}
    \label{Fig: Linear}
\end{figure}

In practice, Eq. \eqref{Equ: Mpo_inverse} is minimized with a tensor-by-tensor strategy, i.e. fixing all local tensors of $\mathcal{U}^{\prime}$ except $\left[B_j\right]$ at site $j$.
The optimization of $\left[B_j\right]$ then reads as in the following
\begin{align}
    \min_{\left[B_j\right]}{\left(e\right)} = \min_{\vec{B_j}}{\left(\vec{B}_j^{\dagger}M_j\vec{B}_j-\vec{B}_j^{\dagger}\vec{N}_j-\vec{N}_j^{\dagger}\vec{B}_j+C\right)}. \hskip 15pt \label{Equ: Linear}
\end{align}
Here we group all the indices of $\left[B_j\right]$ to generate a vector $\vec{B}_j$.
$M_j$ is the environment of $\vec{B}_j$ and $\vec{B}_j^{\dagger}$ in $\textrm{Tr}{\left[\mathcal{U}^{\prime}\mathcal{U}\mathcal{U}^{\dagger}\mathcal{U}^{\prime\dagger}\right]}$, obtained by contracting all tensors in $\textrm{Tr}{\left[\mathcal{U}^{\prime}\mathcal{U}\mathcal{U}^{\dagger}\mathcal{U}^{\prime\dagger}\right]}$ except $\vec{B}_j$ and $\vec{B}_j^{\dagger}$.
$\vec{N}_j$ is the environment of $\vec{B}_j^{\dagger}$ in $\textrm{Tr}{\left[\mathcal{U}^{\dagger}\mathcal{U}^{\prime\dagger}\right]}$ and $C = \textrm{Tr}{\left[\mathds{1}\right]} = d_p^{N_q}$, as shown in Fig. \ref{Fig: Linear}.
With standard contraction strategy for tensor networks, calculation of $M_j$ and $\vec{N}_j$ for each site takes $O(N_q)$ time.
By using caching \cite{Crosswhite2008, Orus2014} one can complete this task in amortized $O(1)$ time.

Now the optimization problem is quadratic in local tensors (a Rayleigh quotient).
The minimization of Eq. \eqref{Equ: Linear} thus corresponds to the solution of the following linear equation
\begin{align}
    M_j\vec{B}_j = \vec{N}_j.
\end{align}
In this sense, the problem of calculating the inverse of a quantum channel is broken into solving linear equations for onsite tensors.
We sweep back and forth with updating the environment tensors $M_j$ and $N_j$ in each iteration step until convergence.
We adopt the canonical form of MPO to accelerate convergence \cite{Noh2020,Cirac2021}.
The time complexity for each iteration is $O(N_q)$.
We denote the convergent $\mathcal{U}^{\prime}$ as $\mathcal{U}^{-1}_{\text{MPO}}$ to approximate $\mathcal{U}^{-1}$.
The convergence criterion is set to $10^{-15}$, which generally can be achieved in ten iterations.

\section{Trace preserving condition}
The inverse of an invertible CPTP map is Hermitian-preserving (HP) and trace-preserving (TP).
These conditions are not imposed on the construction of MPO.
The HP condition requires the Choi matrix to be Hermitian, which is always satisfied during the tensor update process, while the TP condition remains to be verified after taking the inverse and truncation.
The TP condition for a superoperator $\mathcal{U}$ reads $\llangle \mathds{1} | \mathcal{U} = \llangle \mathds{1} |$ \cite{Nielsen2021}, where $\llangle \mathds{1} |$ is the vectorized density operator of the maximally mixed state.
Therefore, we define the trace-infidelity of $\mathcal{U}$ as
\begin{align}
    \overline{F}_{\text{Trace}}(\mathcal{U}) = \left|\llangle \mathds{1} | - \llangle \mathds{1} | \mathcal{U} \right|^{2}.
\end{align}
to describe its deviation from the TP condition.

Generally, if one uses a superoperator $\mathcal{U}$ to approximate a quantum channel $\mathcal{V}$ satisfying the TP condition $\llangle \mathds{1} | \mathcal{V} = \llangle \mathds{1} |$, the trace-infidelity of $\mathcal{U}$ can be estimated as
\begin{align}
    \begin{aligned}
        \overline{F}_{\text{Trace}}(\mathcal{U}) &= \left|\llangle \mathds{1} | - \llangle \mathds{1} | \mathcal{U} \right|^{2} = \left|\llangle \mathds{1} | \mathcal{V} - \llangle \mathds{1} | \mathcal{U} \right|^{2}\\
        &= \llangle \mathds{1} | \left(\mathcal{V} - \mathcal{U}\right)\left(\mathcal{V} - \mathcal{U}\right)^{\dagger} | \mathds{1} \rrangle\\
        &\leq \sum_{n}{\llangle n | \left(\mathcal{V} - \mathcal{U}\right)\left(\mathcal{V} - \mathcal{U}\right)^{\dagger} | n \rrangle}\\
        &= \textrm{Tr}{\left[\left(\mathcal{V} - \mathcal{U}\right)\left(\mathcal{V} - \mathcal{U}\right)^{\dagger}\right]}\\
        &= \llvert\mathcal{V} - \mathcal{U}\rrvert_F^2,
    \end{aligned}
\end{align}
where we use a set of basis $\{|n\rrangle\}$ including $|\mathds{1}\rrangle$ to expand the trace.
The inequality is due to the semidefinite positivity of quadratic form $\left(\mathcal{V} - \mathcal{U}\right)\left(\mathcal{V} - \mathcal{U}\right)^{\dagger}$.
Therefore, as $\llvert\mathcal{V} - \mathcal{U}\rrvert_F$ decreases, which is just (or equivalent to) what we try to minimize when taking the inverse or truncation, $\overline{F}_{\textrm{Trace}}$ will simultaneously decrease and the TP condition will be satisfied automatically.
In Appendix \ref{App: PEPO}, we numerically verify this property by monitoring the change of the trace-infidelity during the iteration process when taking the inverse of a 2D noisy quantum circuit represented by PEPO.

\section{Implementation of MPO-based quantum error mitigation}
In summary, our MPO-based QEM consists of five steps.

1. Decomposition of the entire circuit.
We divide the noisy circuit to be mitigated into $m$ parts, each with $d_0$ layers.
For each part, we implement the following steps.

2. Implementation of quantum process tomography.
We apply QPT on one part of the noisy quantum circuit $\mathcal{U}_{[k]}$ to determine its MPO representation $\mathcal{U}_{\textrm{MPO}, [k]}$, as shown in Fig. \ref{Fig: MPO_QEM}(a).
The QPT method introduced by Torlai \textit{et al.} \cite{Torlai2020} (see Appendix \ref{App: QPT}) can be used to parameterize a noisy quantum circuit with LPDO via unsupervised learning.

3. Calculation of the circuit inverse.
We employ the MPO-inverse method to represent the inverse of $\mathcal{U}_{[k]}$ as an MPO $\mathcal{U}^{-1}_{\textrm{MPO}, [k]}$, as shown in Fig. \ref{Fig: MPO_QEM}(b).
We assume the quantum channel $\mathcal{U}_{[k]}$ is invertible, which is generally satisfied in practice.

4. Calculation of the noise inverse.
We calculate the total effect of all errors in the $k$-th part by contracting the ideal quantum circuit $\mathcal{U}^{\left(0\right)}_{[k]}$ with the inverse of noisy circuit $\mathcal{U}^{-1}_{\textrm{MPO}, [k]}$.
The contraction strategy is similar to the evolution of MPS \cite{Verstraete2008, Orus2014}, i.e., contracting the circuit and truncating the resulting MPO layer by layer \cite{Noh2020, Zhou2020}.
In the end, we obtain the MPO representation of the inverse noise channel $\mathcal{E}^{-1}_{\textrm{MPO}, {[k]}}$ shown in Fig. \ref{Fig: MPO_QEM}(c).

5. Compensation for the errors.
We construct quantum circuits that simulate the map $\mathcal{E}^{-1}_{\textrm{MPO}, [k]}$, which can be decomposed into linear combinations of experimentally accessible CPTP maps \cite{Jiang2021}.
In our method, to avoid extra errors induced by the correcting circuit simulating $\mathcal{E}^{-1}_{\textrm{MPO}, [k]}$, we only keep $D^{\prime} = 1$ and apply the corresponding single-qubit operations after $\mathcal{U}_{[k]}$ to mitigate the errors in this part.

\section{Numerical simulations}
\subsection{Calculation of the circuit inverse $\mathcal{U}^{-1}_{\rm MPO}$}
One of the key points of our method lies in the second step of the whole process, i.e., whether $\mathcal{U}^{-1}$ can be approximated as an MPO with a similar bond dimension to $\mathcal{U}$ or not.
It is not satisfied by a general MPO but is a useful property for a noisy circuit $\mathcal{U}$ that slightly deviates from the ideal circuit $\mathcal{U}^{(0)}$, since for the latter, taking the inverse is equivalent to taking the Hermitian conjugate, meaning that ${\mathcal{U}^{(0)}}^{-1} = {\mathcal{U}^{(0)}}^{\dagger}$ and $\mathcal{U}^{(0)}$ share rigorously the same bond dimension in their MPO representations.

\begin{figure}
    \centering
    \includegraphics[width = 0.85\linewidth]{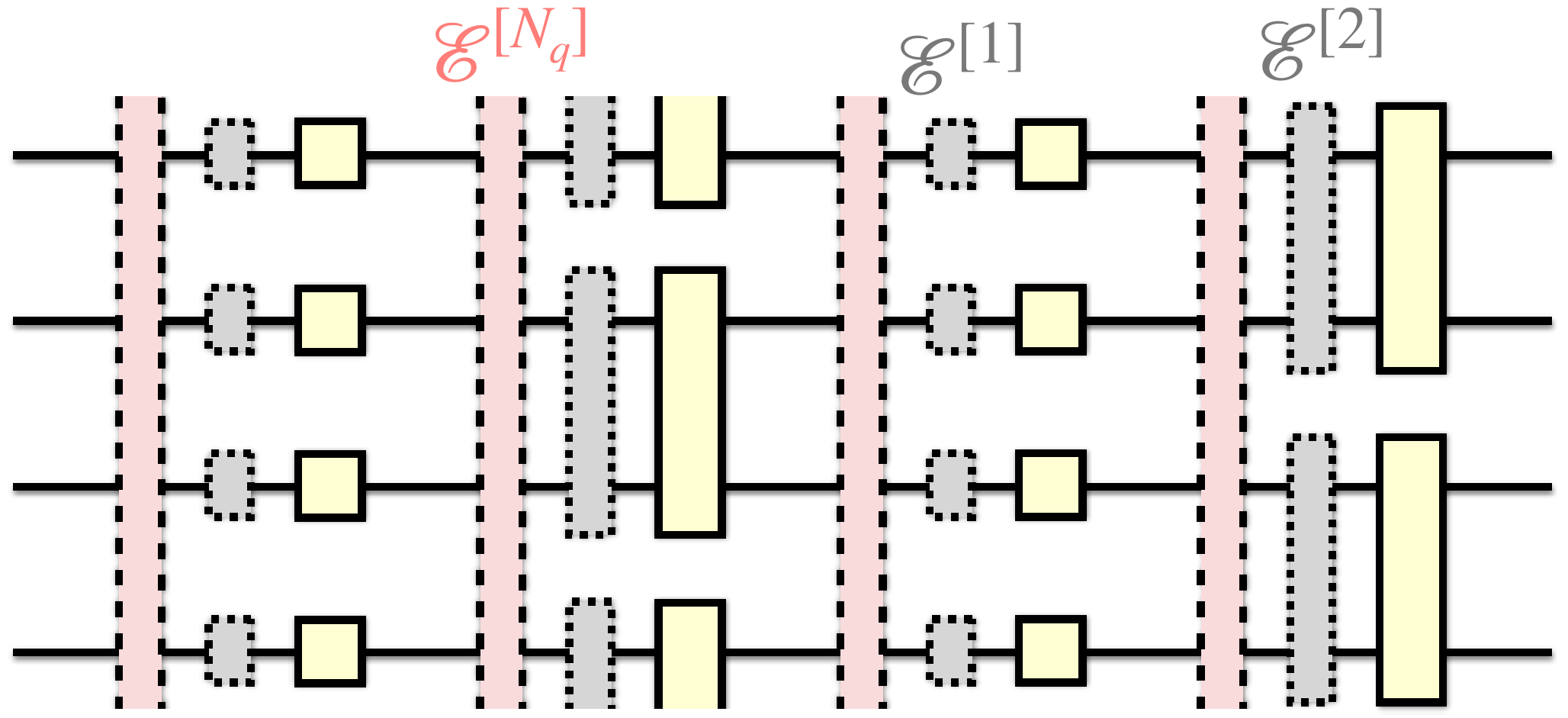}
    \caption{(Color online) The test circuit configuration with $\rm{depth}=4$ and $N_q=4$.
    We add $i$-qubit ($i=1, 2$) noise $\mathcal{E}^{[i]}$ after each ideal $i$-qubit gate and global $N_q$-qubit depolarizing noise $\mathcal{E}^{[N_q]}$ after each layer.}
    \label{Fig: Testcircuit}
\end{figure}

We first test our MPO-inverse method on noisy quantum circuits, which is the second step of our QEM approach shown in Fig. \ref{Fig: MPO_QEM}(b).
We will make use of the test circuit shown in Fig. \ref{Fig: Testcircuit} with $\rm{depth} = 4$, as commonly adopted in QPT \cite{Torlai2020} and QEM \cite{Temme2017}.
In this circuit, the odd layer is a tensor product of $N_q/2$ two-qubit CNOT gates, while the even layer is a tensor product of $N_q$ single-qubit gates randomly chosen from $\{Z, H, S, T\}$.
This circuit has an exact MPO representation with $D = 4$.
We set $N_q = 10$ and introduce noise channels after each gate, including depolarizing noise, dephasing noise, bit flipping noise, and amplitude damping noise.
The error rate for each $i$-qubit gate ($i = 1, 2$) is randomly chosen from $[0.8\varepsilon_i, 1.2\varepsilon_i]$, where $\varepsilon_i$ is denoted as the average error rate with fixing $\varepsilon_2 = 10\varepsilon_1$ hereafter.
We provide more details on numerical settings in Appendix \ref{App: Numeric}.
\begin{figure*}
    \centering
    \includegraphics[width=\linewidth]{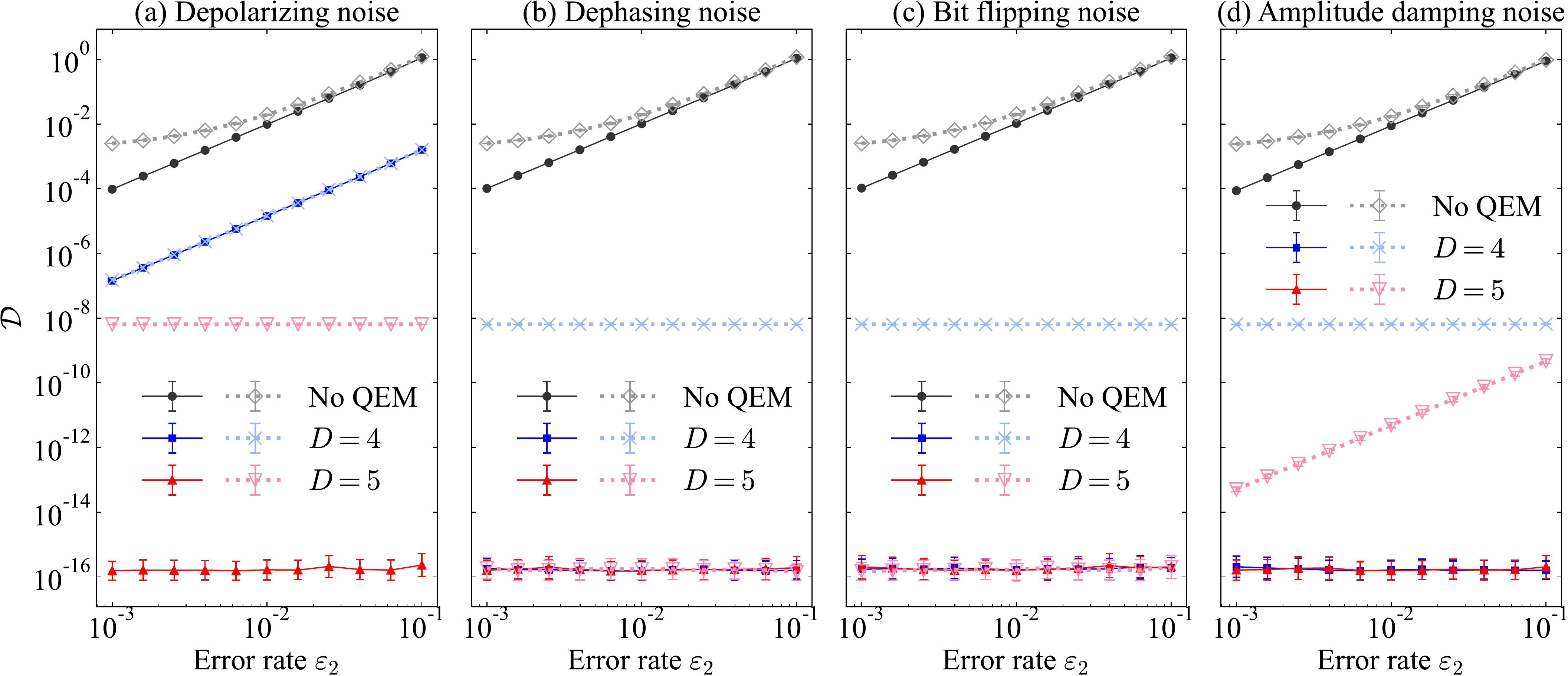}
    \caption{(Color online) Performance of our MPO-inverse method.
    Solid lines and dotted lines describe circuits without and with global $N_q$-qubit depolarizing noise $\mathcal{E}^{[N_q]}$ respectively, the error rate of which is $\varepsilon_n = 0.01$.
    The system size is $N_q = 10$.
    We plot $\mathcal{D}(\mathcal{U}^{-1}_{\rm{MPO}}\circ\mathcal{U}, \mathds{1})$ for different average error rates and benchmark $\mathcal{D}(\mathcal{U}, \mathcal{U}^{(0)})$ for the total noise effect in the original circuit, labeled as ``No QEM''.}
    \label{Fig: Mpo_invserse}
\end{figure*}

We employ our MPO-inverse method to calculate the inverse of the real circuit $\mathcal{U}^{-1}_{\textrm{MPO}}$ for $D = 4\text{ and }5$, then evaluate the relative distance between $\mathcal{U}^{-1}_{\rm{MPO}}\circ\mathcal{U}$ and identity $\mathds{1}$ in Fig. \ref{Fig: Mpo_invserse} (solid lines).
The relative distance between two superoperators $\mathcal{U}$ and $\mathcal{V}$ is defined as $\mathcal{D}\left(\mathcal{U}, \mathcal{V}\right) = \llvert\mathcal{U} - \mathcal{V}\rrvert_F^2/\sqrt{\llvert\mathcal{U}\rrvert_F^2\llvert\mathcal{V}\rrvert_F^2}$, which bounds the distance of output density matrices and can be directly estimated via MPO contraction.
The result is compared with $\mathcal{D}\left(\mathcal{U}^{(0)-1}\circ \mathcal{U}, \mathds{1}\right) = \mathcal{D}\left(\mathcal{U}, \mathcal{U}^{(0)}\right)$ (since $\mathcal{U}^{(0)}$ is unitary), which represents the total noise effect.
It is demonstrated that with $D=4$, the error induced by taking the inverse is several orders of magnitude smaller than the circuit error itself.
Moreover, with $D=5$, i.e., with just one extra virtual bond dimension compared to $\mathcal{U}^{(0)}$ (and thus ${\mathcal{U}^{(0)}}^{-1}$), the accuracy of the MPO representation $\mathcal{U}^{-1}_{\rm MPO}$ is near-perfect for all four types of noise.

We next move to a much more non-trivial case, where we add extra global $N_q$-qubit depolarizing noise after each layer, which can be written as an MPO with $D = 2$.
The global depolarizing noise is a standard model to approximate the decoherence process of the entire circuit \cite{Dur2005}, which has drawn the attention of experimentalists when applying QEM to real devices \cite{Vovrosh2021, Urbanek2021} but cannot be mitigated with the standard each-gate QEM approach.
We fix the $N_q$-qubit error rate at $\varepsilon_n = 10^{-2}$ and repeat the previous calculations, as shown by dotted lines in Fig. \ref{Fig: Mpo_invserse}.
With $D = 5$, the accuracy of MPO-inverse can still achieve $10^{-8}$.
The result also implies that the circuit with depolarizing noise is harder to handle than the other three types of noise.
In summary, one can capture almost all the noise effects during the procedure of MPO-inverse with just one extra virtual bond dimension compared to ${\mathcal{U}^{(0)}}^{-1}$ to represent $\mathcal{U}^{-1}_{\textrm{MPO}}$.

\subsection{Calculation of the noise inverse $\mathcal{E}^{-1}_{\rm MPO}$}
\begin{figure}
    \centering
    \includegraphics[width = \linewidth, height = 0.75\linewidth]{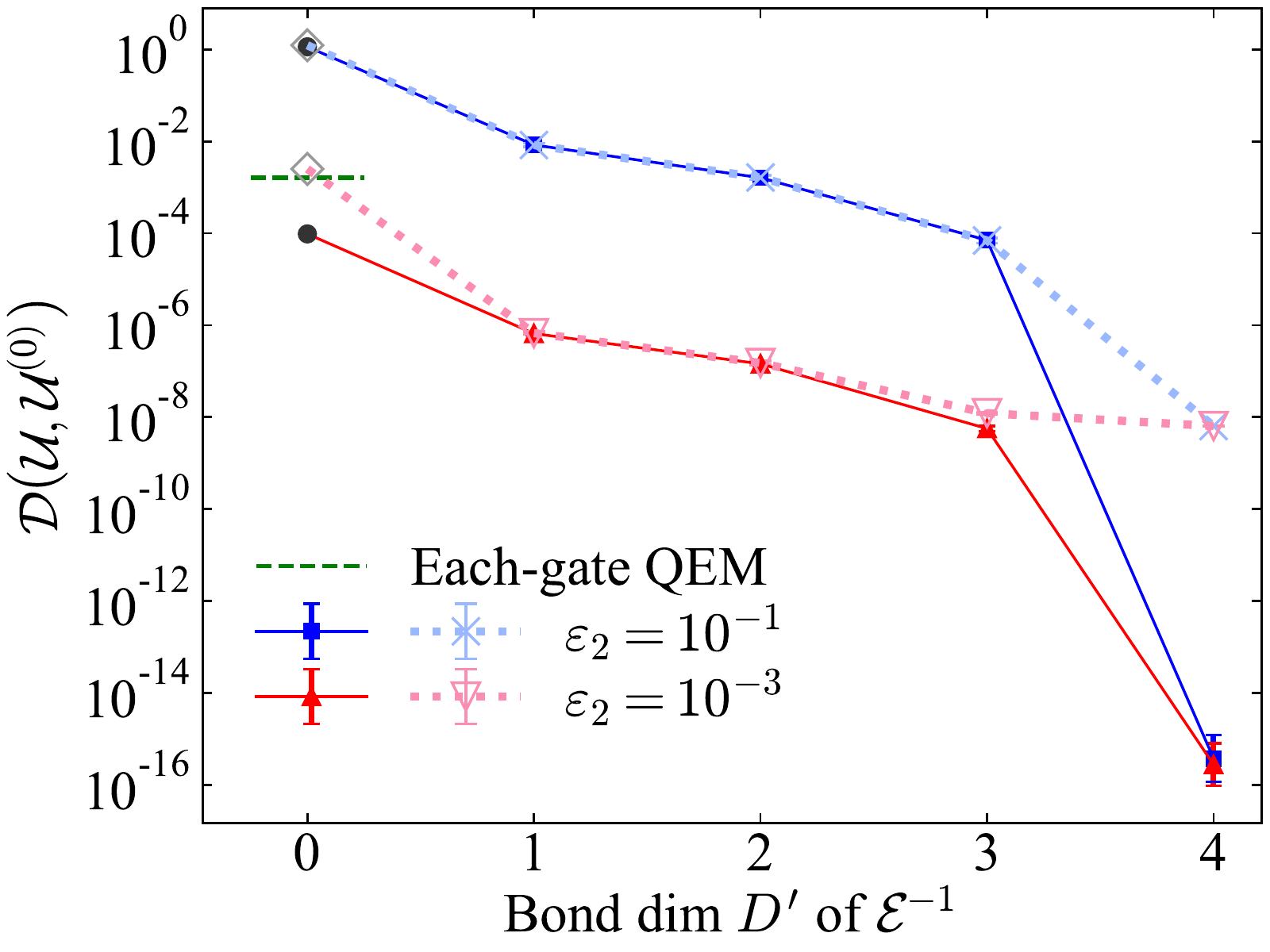}
    \caption{(Color online) $\mathcal{D}(\mathcal{E}^{-1}_{\rm{MPO}}\circ\mathcal{U}, \mathcal{U}^{(0)})$ with different bond dimension $D^{\prime}$ for $\mathcal{E}^{-1}_{\rm{MPO}}$.
    Here $D^{\prime} = 0$ corresponds to $\mathcal{D}(\mathcal{U}, \mathcal{U}^{(0)})$.
    Solid lines and dotted lines describe circuits without and with global $N_q$-qubit depolarizing noise $\mathcal{E}^{[N_q]}$ respectively, the error rate of which is $\varepsilon_n = 0.01$.
    The system size is $N_q = 10$.
    For circuits with global noise, we plot the bound of QEM performance if one mitigates the error of each gate independently, labeled as ``Each-gate QEM''.}
    \label{Fig: Noise_inverse}
\end{figure}

We then explore the power of MPO to represent the noise inverse shown in Fig. \ref{Fig: MPO_QEM}(c) on the same noisy test circuit of the previous example with $N_q = 10$ and $\rm{depth}=4$.
We focus on the hardest case, where the depolarizing noise is applied after each gate.
We start with a $D = 5$ MPO representation of the circuit inverse $\mathcal{U}^{-1}_{\rm{MPO}}$ obtained above, then contract $\mathcal{U}^{(0)}$ with $\mathcal{U}^{-1}_{\rm{MPO}}$ to calculate the inverse of the noise channel $\mathcal{E}^{-1}_{\rm{MPO}}$ with the bond dimension $D^{\prime}$.
In experiments, one needs to implement $\mathcal{E}^{-1}_{\rm{MPO}}$ with real quantum circuits.
Therefore, a $D^{\prime}$ as small as possible is desired to reduce experimental resources and avoid extra errors, meaning that we need to truncate it when calculating $\mathcal{E}^{-1}_{\rm{MPO}}$.
We evaluate $\mathcal{D}\left(\mathcal{E}^{-1}_{\rm{MPO}}\circ \mathcal{U}, \mathcal{U}^{\left(0\right)}\right)$ for different $D^{\prime}$ at two average error rates $\varepsilon_2 = 10^{-1}$ and $\varepsilon_2 = 10^{-3}$ in Fig. \ref{Fig: Noise_inverse} without (solid lines) and with (dotted lines) global depolarizing noise, where $D^{\prime}=0$ corresponds to $\mathcal{D}(\mathcal{U}, \mathcal{U}^{(0)})$, i.e., the lines labeled as ``No QEM'' in Fig. \ref{Fig: Mpo_invserse}(a).
It is shown that with $D^{\prime} = 1$, we can suppress the noise effect by two orders of magnitude, while for $D^{\prime}=4$, we can achieve the same accuracy as the MPO-inverse step, i.e., the red lines in Fig. \ref{Fig: Mpo_invserse} (a).

Moreover, after applying $\mathcal{E}^{-1}_{\textrm{MPO}}$, the final accuracy with and without global noise for the same bond dimension $D^{\prime}$ close to each other when $D^{\prime} = 1, 2, 3$, indicating the power of our method in characterizing correlated noises.
On the contrary, if one tries to mitigate the error of every gate independently without accounting for these correlated errors, the accuracy of characterizing the noise channel and hence the performance of QEM are limited \cite{Cao2021}.
To estimate the bound, we assume that all local errors in the circuit are perfectly mitigated while non-local errors remain, and the resulting distance between the noisy circuit after QEM and the ideal circuit is shown by the dashed line in Fig. \ref{Fig: Noise_inverse}.
We conclude that MPO serves as a good representation for a variety of noise channels, capable of characterizing and mitigating almost all the errors in the circuit, including non-local ones.

\subsection{Simulation of QEM on deep circuits}\label{Sec: Deep}
In this subsection, we simulate our full QEM approach on noisy circuits with deeper depth.
The gate configuration in our test circuit is the same as that in previous sections, with the circuit size being $N_q = 20$ and $\textrm{depth}=20$.
The noise channel added after each ideal gate is randomly chosen from the four noise models we have considered above.
The average error rate for two-qubit gates is $\varepsilon_2=10^{-2}$, while that for single-qubit gates is one order of magnitude smaller, i.e., $\varepsilon_1 = \varepsilon_2 / 10 = 10^{-3}$.
We divide the circuit into $m = 5$ parts, each with $d_0 = 4$ layers, and add global $N_q$-qubit depolarizing noise after each part to simulate global decoherence of the system induced by, e.g. coupling with the environment during our implementation of QEM, whose error rate is fixed at $\varepsilon_n = 0.05$.

For the first step, we simply use the standard contraction and truncation method \cite{Verstraete2008, Orus2014} to obtain an MPO approximation $\mathcal{U}_{\rm{MPO}, [k]}$ of the $k$-th part of the real circuit with $D = 5$.
Next, we employ our MPO-inverse method and truncation method to obtain the MPO representation of the noise inverse $\mathcal{E}^{-1}_{\textrm{MPO}, [k]}$ with $D^{\prime} = 1$.
Then we insert the corresponding single-qubit operations realizing $\mathcal{E}^{-1}_{{\textrm{MPO}}, [k]}$ into the end of the $k$-th part, which is assumed to be noisy themselves and thus each gate is followed by a single-qubit depolarizing noise channel with the average error rate $\varepsilon_1 = 10^{-3}$.

\begin{figure}
    \centering
    \includegraphics[width = \linewidth, height = 0.75\linewidth]{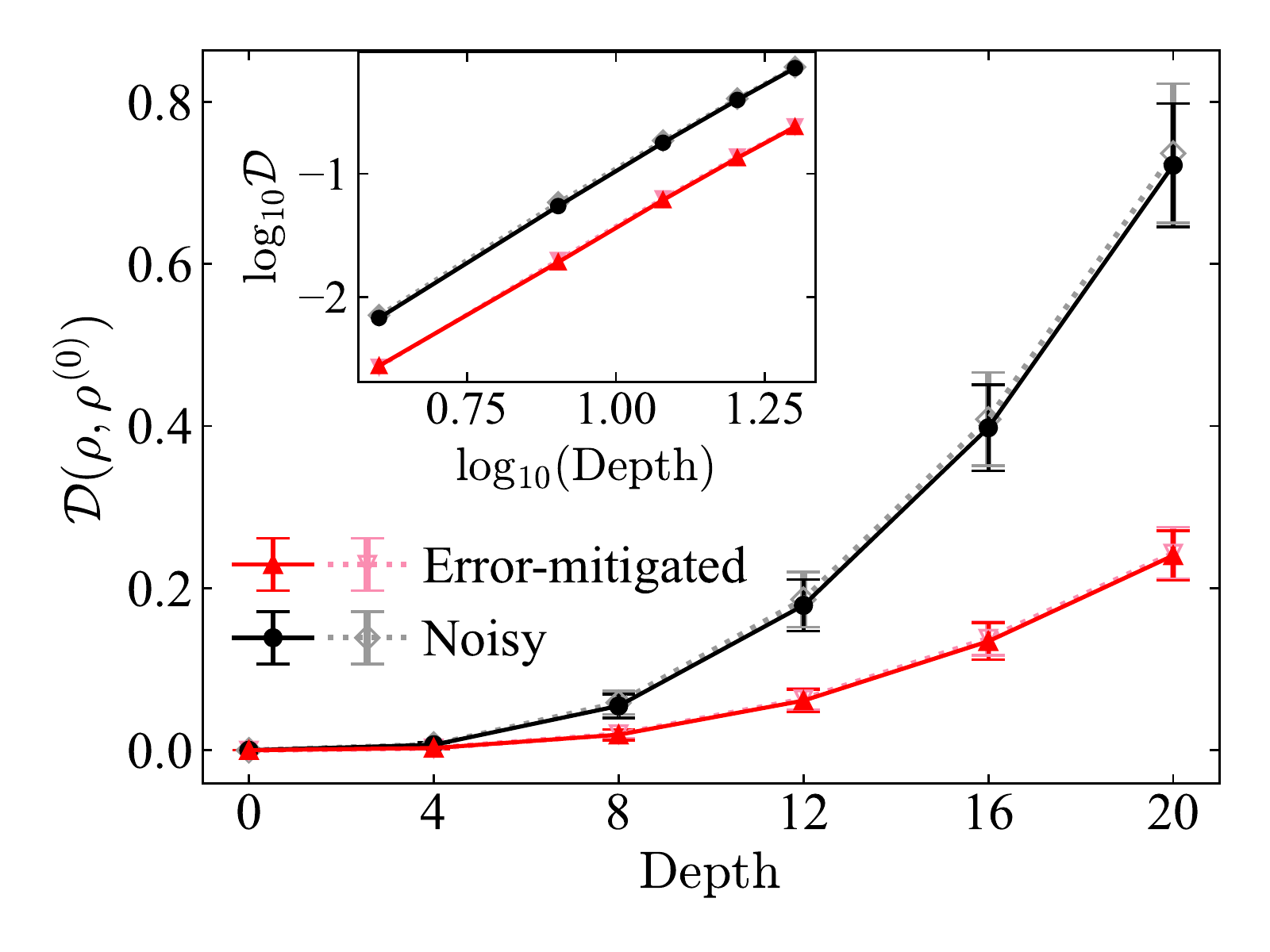}
    \caption{(Color online) Distance between $\rho$ from noisy or error-mitigated circuits and $\rho^{(0)}$ from ideal circuits.
    The average error rate for two-qubit gates is $\varepsilon_2 = 0.01$, and the system size is $N_q = 20$.
    Solid lines and dotted lines describe circuits without and with global $N_q$-qubit depolarizing noise $\mathcal{E}^{[N_q]}$ respectively, the error rate of which is $\varepsilon_n = 0.05$.
    We take the logarithmic in the inset.}
    \label{Fig: QEM}
\end{figure}

Experimentally valuable measurements include single-qubit observables such as magnetism and two-qubit correlation functions.
However, to demonstrate the performance of our QEM method, we directly reconstruct the output density matrices, which is a more challenging task than recovering the ideal measurement results.
The input state is chosen as the product state $\otimes_{i=1}^{N_q}\ket{0}_i$.
We evaluate the relative distance between the output density matrices $\rho$ reconstructed from error-mitigated circuits and ideal output $\rho^{(0)}$ in Fig. \ref{Fig: QEM}, which is defined as $\mathcal{D}\left(\rho, \rho^{(0)}\right) = \llvert \rho-\rho^{(0)}\rrvert_F^2 / \sqrt{\llvert \rho\rrvert_F^2\llvert \rho^{(0)}\rrvert_F^2}$.
It is demonstrated that with the correcting circuits, though modeled as noisy themselves, the noise effects in the output states can be suppressed by about a factor of three, which is robust against the existence of global noise.

In addition, we plot the logarithmic distance in the inset of Fig. \ref{Fig: QEM} to study the accumulation of errors in deep circuits, which shows a power law of the output distance with respect to the circuit depth, i.e.,
\begin{align}
    \mathcal{D}\left(\rho, \rho^{(0)}\right) \propto \textrm{depth}^{\alpha}.\label{equ: deep}
\end{align}
Fitting from the noisy and error-mitigated data gives approximately the same exponent $\alpha \approx 2.8$.
In other words, errors in the output states are suppressed by a constant factor $10^{\Delta b}\approx 2.9$ independent of the circuit depth, where $\Delta b$ is the difference of intercepts of the red and black lines in the inset, fitted as $\Delta b \approx 0.46$.
This implies that our method is scalable with the circuit depth. 

\begin{figure}
    \centering
    \includegraphics[width = \linewidth]{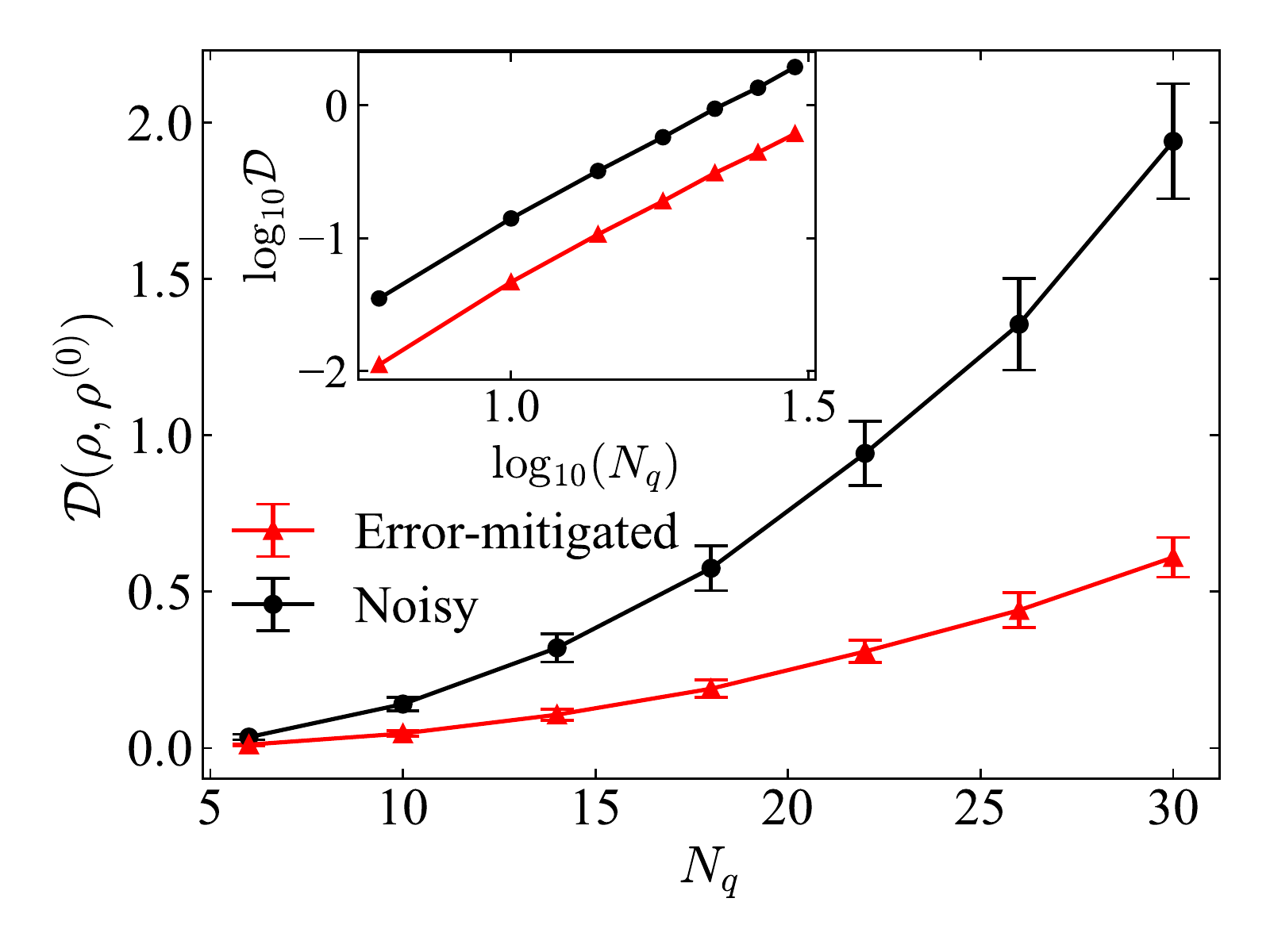}
    \caption{(Color online) Distance between $\rho$ from noisy or error-mitigated circuits without global noise and $\rho^{(0)}$ from ideal circuits with $\rm{depth}=20$.
    The average error rate for two-qubit gates is $\varepsilon_2 = 0.01$ and we take the logarithmic in the inset.}
    \label{Fig: QEM_N}
\end{figure}

To study the influence of the system size $N_q$ on the above results, we further test our method on circuits with the same gate and noise configuration but different $N_q$, as shown in Fig. \ref{Fig: QEM_N}.
We only focus on circuits with $\rm{depth}=20$ undergoing local noise to benchmark.
The relation in Eq. \eqref{equ: deep} can be extended by adding the dependence on $N_q$ as
\begin{align}
    \mathcal{D}\left(\rho, \rho^{(0)}\right) \propto \textrm{depth}^{\alpha}N_q^{\beta},
\end{align}
where $\beta\approx 2.5$ is given by linear fitting from the error-mitigated data (red lines) in the inset of Fig. \ref{Fig: QEM_N}.
Moreover, the red and black lines in the inset are parallel, which indicates that the circuit size $N_q$ is irrelevant to the performance of our QEM method, suggesting its application to larger systems.
However, the exponent $\alpha$ may slightly vary for different $N_q$, which is carefully studied in Appendix \ref{App: Finite_size}.
In summary, our method is scalable with both the circuit depth and the system size, at least for these accessible quantum devices in the NISQ era.

\begin{figure*}
    \centering
    \includegraphics[width = \linewidth]{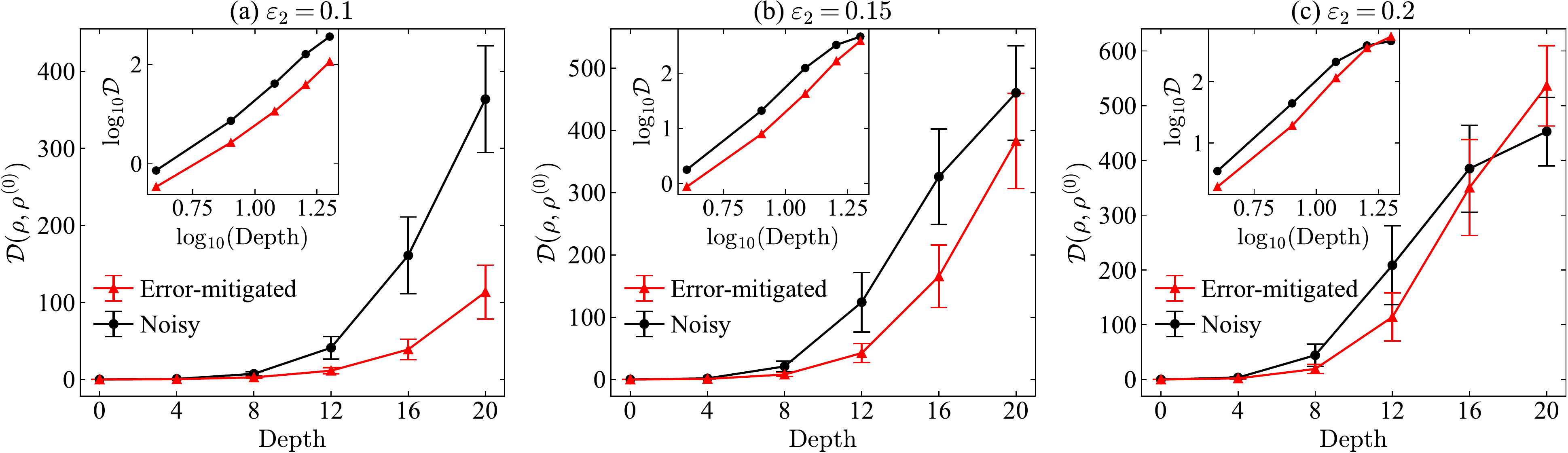}
    \caption{(Color online) Distance between $\rho$ from noisy or error-mitigated circuits without global noise and $\rho^{(0)}$ from ideal circuits for different $\varepsilon_2$.
    The system size is $N_q = 20$ and we take the logarithmic in the inset.}
    \label{Fig: Diff_err}
\end{figure*}

Similar to other error correction or mitigation methods, our MPO-based QEM scheme also has a threshold of the error rate, since we have assumed that the noisy circuit deviates slightly from the ideal one to promise an efficient MPO representation of the noise inverse $\mathcal{E}^{-1}_{\rm MPO}$.
We test our method for three large average error rates of two-qubit gates $\varepsilon_2$ in Fig. \ref{Fig: Diff_err} with $N_q = 20$, which implies the existence of a threshold at $\varepsilon_2\approx 0.15$, far higher than the benchmarks in present-day quantum devices.

All numerical simulations are repeated 200 times, where the circuit and noise configuration are sampled independently for each simulation.
The geometric means and geometric standard deviations are plotted in Fig. \ref{Fig: Mpo_invserse} and \ref{Fig: Noise_inverse}, while in Fig. \ref{Fig: QEM}, \ref{Fig: QEM_N}, and \ref{Fig: Diff_err} we adopt the arithmetic means and standard deviations.

\section{Conclusions and discussions}
We introduce a variational algorithm to calculate the inverse of a noisy quantum circuit with matrix product operators.
The validity of our method is established using noisy quantum circuits undergoing four types of noise with error rates up to $10^{-1}$.

In addition, we propose a quantum error mitigation scheme based on the MPO representation of the noise channel.
Our QEM method is tested using numerical simulations on noisy quantum circuits with global depolarizing noise.
We show that with only a small bond dimension $D^{\prime} = 1$ for the inverse noise channel $\mathcal{E}^{-1}$, which is simply a direct product of single-qubit maps on each site, the total noise in the reconstructed density operator is suppressed by several times, despite the correcting circuits themselves being noisy.

Compared with other QEM techniques proposed in recent years, our method treats all errors within several layers as a whole noise channel via MPO representation, including long-range spatially correlated and short-range temporally correlated ones.
In this sense, one can characterize the noise channel more accurately, improving the QEM performance \cite{Cao2021}.

With the parameterization of quantum channels via tensor networks, modeling the noise channel is scalable with system size $N_q$ and can be accomplished with polynomial overhead.
Moreover, the errors in output states are suppressed by a constant factor regardless of the circuit depth, which shows the scalability of our method for deeper circuits in the NISQ era.

Some interesting results in our work are worth further study.
For example, what is the difference between these noise models considered in Fig. \ref{Fig: Mpo_invserse}?
Why some lines are parallel to the error-unmitigated data (the black lines) indicating the remaining errors to be local ones, while others are constants independent of the two-qubit error rate $\varepsilon_2$, which implies the residue of global noise?
We believe that with the MPO representations of noise channels, these issues will be fully understood and addressed, which may inspire specialists in the community of quantum information to study the decoherence effects of quantum noise from the entanglement perspective \cite{Guo2022}.

The generalization of our method to circuits in higher spatial dimensions is hopeful.
With the help of projected entangled pair operators (PEPO), one may characterize and mitigate 2D noise channels.
Our MPO-inverse method can be naturally generalized to PEPO, which is verified on 2D noisy quantum circuits in Appendix \ref{App: PEPO}.

We anticipate that our QEM method can be implemented on larger and deeper quantum devices with many more qubits and state-of-the-art hardware error rates.
It will enable medium-sized quantum computers, on which quantum error correction codes are hard to realize, to carry out complicated quantum algorithms or quantum-classical hybrid algorithms with high fidelity.

\begin{acknowledgments}
    This work is supported by the National Natural Science Foundation of China (NSFC) (Grant No. 12174214 and No. 92065205), the National Key R\&D Program of China (Grant No. 2018YFA0306504), and the Innovation Program for Quantum Science and Technology (Project 2-9-4).
\end{acknowledgments}

\appendix
\renewcommand{\theequation}{S\arabic{equation}} \setcounter{equation}{0}
\renewcommand{\thefigure}{S\arabic{figure}} \setcounter{figure}{0}


\section{Quantum process tomography method with tensor networks}\label{App: QPT}
The first step of our QEM framework is to obtain the MPO representation of noisy quantum circuits using the QPT method proposed by Torlai \textit{et al.} \cite{Torlai2020}.
They use LPDO to represent the Choi matrix of a quantum channel and update the tensors via unsupervised learning.

For an unknown quantum channel $\mathcal{U}$, product states are prepared as input states $\bm{\rho}_{\bm{i}} = \otimes_{k=1}^{N_q}\rho_{i_k}$, while positive operator valued measurements (POVM) $\bm{M}_{\bm{j}} = \otimes_{k=1}^{N_q}M_{j_k}$ are applied after the quantum channel.
Experimental results are converted into a conditional probability distribution of strings $\bm{i} = (i_1, \dots, i_{N_q})$ and $\bm{j} = (j_1, \dots, j_{N_q})$, denoted as $P_{\mathcal{U}}\left(\bm{j}|\bm{i}\right)$.

To reconstruct the quantum channel $\mathcal{U}$, one updates tensors of the LPDO representing a quantum channel $\mathcal{V}$ with the gradient descent method to minimize the distance between two probability distributions.
It is commonly characterized by the Kullbach-Leibler (KL) divergence
\begin{align}
    \begin{aligned}
        D_{\text{KL}} &= \sum_{\bm{j}} P_{\mathcal{U}}\left(\bm{j}\right) \log{\frac{P_{\mathcal{U}}\left(\bm{j}\right)}{P_{\mathcal{V}}\left(\bm{j}\right)}}\\
        &= \sum_{\bm{i}, \bm{j}} P\left(\bm{i}\right) P_{\mathcal{U}}\left(\bm{j}|\bm{i}\right)\log{\frac{P_{\mathcal{U}}\left(\bm{j}|\bm{i}\right)}{P_{\mathcal{V}}\left(\bm{j}|\bm{i}\right)}}
    \end{aligned}
\end{align}
where $P\left(\bm{i}\right)$ is the prior distribution of input states. 
$P_{\mathcal{U}}\left(\bm{j}|\bm{i}\right)$ is estimated through measurements, while $P_{\mathcal{V}}\left(\bm{j}|\bm{i}\right)$ is the corresponding distribution simulated from $\mathcal{V}$, i.e.,
\begin{align}
    P_{\mathcal{V}}\left(\bm{j}|\bm{i}\right)=\textrm{Tr}_{\bm{\sigma}, \bm{\tau}}\left[\left(\bm{\rho}_{\bm{i}}^T\otimes \bm{M}_{\bm{j}}\right)\Lambda_{\mathcal{V}}\right]
\end{align}
The cost function and its gradient can be calculated efficiently via standard tensor contraction.

In this method, the total number of parameters to represent a quantum channel scales linearly with the system size $N_q$, and so does the required number of state preparations and measurements, as verified numerically in their study.

\section{Settings of numerical simulations}\label{App: Numeric}
\begin{figure*}
    \centering
    \includegraphics[width = 0.7\linewidth]{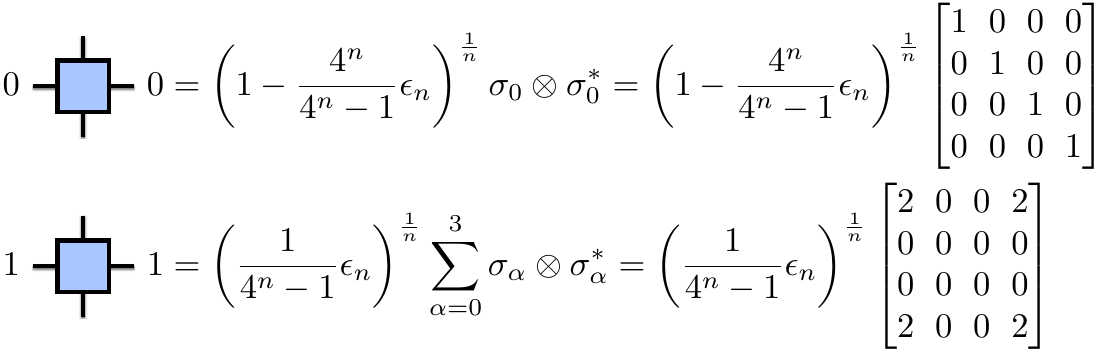}
    \caption{(Color online) Nonzero tensor elements in the MPO representation of depolarizing noise.}
    \label{Fig: Dep_MPO}
\end{figure*}

In our 1D test circuit, the odd layer is a tensor product of $N_q/2$ two-qubit controlled-NOT (CX) gates
\begin{align}
    \text{CX} = \begin{bmatrix}
    1 & 0 & 0 & 0 \\
    0 & 1 & 0 & 0 \\
    0 & 0 & 0 & 1 \\
    0 & 0 & 1 & 0 \\
    \end{bmatrix}\:,
\end{align}
while the even layer is a tensor product of $N_q$ single-qubit gates randomly chosen from four commonly used gates in quantum computation, including the $Z$ gate
\begin{align}
    Z=\begin{bmatrix}
    1 & 0\\
    0 & -1
    \end{bmatrix}\:,
\end{align}
the Hadamard gate
\begin{align}
    H=\frac{1}{\sqrt{2}}\begin{bmatrix}
    1 & 1\\
    1 & -1
    \end{bmatrix}\:,
\end{align}
the phase gate
\begin{align}
    S=\begin{bmatrix}
    1 & 0\\
    0 & i
    \end{bmatrix}\:,
\end{align}
and the $\pi/8$ gate
\begin{align}
    T=\begin{bmatrix}
    1 & 0\\
    0 & \exp{\left(\frac{\pi}{4}i\right)}
    \end{bmatrix}\:.
\end{align}
Superoperators of these unitary gates can be directly constructed, similar to Eq. \eqref{equ: super}.

In numerical simulations for noisy circuits, the noise added after each gate includes four types.
The depolarizing noise is defined as
\begin{align}
    \mathcal{E}^{[1]}\left(\rho^{[1]}\right) = \left(1-\frac{4}{3}\epsilon_1\right)\rho^{[1]} + \frac{1}{3}\epsilon_1\sum_{i=0}^{3}\sigma_i\rho^{[1]}\sigma_i
\end{align}
for a single-qubit state $\rho^{[1]}$, and
\begin{align}
    \begin{aligned}
        \mathcal{E}^{[2]}\left(\rho^{[2]}\right) &= \left(1-\frac{16}{15}\epsilon_2\right)\rho^{[2]}\\
        &+ \frac{1}{15}\epsilon_2\sum_{i,j=0}^{3}\left(\sigma_i\otimes\sigma_j\right)\rho^{[2]}\left(\sigma_i\otimes\sigma_j\right)
    \end{aligned}
\end{align}
for a two-qubit state $\rho^{[2]}$.
The dephasing noise is defined as
\begin{align}
    \mathcal{E}^{[1]}\left(\rho^{[1]}\right) = \left(1-2\epsilon_1\right)\rho^{[1]} + \epsilon_1\sum_{i\in\{0, 3\}}\sigma_i\rho^{[1]}\sigma_i
\end{align}
for a single-qubit state $\rho^{[1]}$, and
\begin{align}
    \begin{aligned}
        \mathcal{E}^{[2]}\left(\rho^{[2]}\right) &= \left(1-\frac{4}{3}\epsilon_2\right)\rho^{[2]}\\
        &+ \frac{1}{3}\epsilon_2\sum_{i,j\in\{0, 3\}}\left(\sigma_i\otimes\sigma_j\right)\rho^{[2]}\left(\sigma_i\otimes\sigma_j\right)
    \end{aligned}
\end{align}
for a two-qubit state $\rho^{[2]}$.
The bit flipping noise is defined as
\begin{align}
    \mathcal{E}^{[1]}\left(\rho^{[1]}\right) = \left(1-\epsilon_1\right)\rho^{[1]} + \epsilon_1\sigma_x\rho^{[1]}\sigma_x
\end{align}
for a single-qubit state $\rho^{[1]}$, and
\begin{align}
    \begin{aligned}
        \mathcal{E}^{[2]}\left(\rho^{[2]}\right) = \left(1-\epsilon_2\right)\rho^{[2]} + \epsilon_2\left(\sigma_x\otimes\sigma_x\right)\rho^{[2]}\left(\sigma_x\otimes\sigma_x\right)
    \end{aligned}
\end{align}
for a two-qubit state $\rho^{[2]}$.
The amplitude damping noise is defined by the Kraus operator $E_0 = \ket{0}\hspace{-1mm}\bra{0} + \sqrt{1-\varepsilon}\ket{1}\hspace{-1mm}\bra{1}$ and $E_1 = \sqrt{\varepsilon}\ket{0}\hspace{-1mm}\bra{1}$ with the operator-sum representation
\begin{align}
    \mathcal{E}^{[1]}\left(\rho^{[1]}\right) = E_0\rho^{[1]} E_0^{\dagger} + E_1\rho^{[1]} E_1^{\dagger}.
\end{align}
for a single-qubit state $\rho^{[1]}$, and
\begin{align}
    \mathcal{E}^{[2]}\left(\rho^{[2]}\right) = \mathcal{E}^{[1]}_1 \otimes \mathcal{E}^{[1]}_2\left(\rho^{[2]}\right)
\end{align}
for a two-qubit state $\rho^{[2]}$.
The error rate $\epsilon_i$ for each i-qubit gate is randomly chosen from $[0.8\varepsilon_i, 1.2\varepsilon_i]$, while $\varepsilon_i$ is denoted as the average error rate in the main text.

The global $n$-qubit depolarizing noise with the error rate $\epsilon_n$ is defined as
\begin{align}
    \begin{aligned}
        \mathcal{E}^{[n]}\left(\rho^{[n]}\right) &= \left(1-\frac{4^n}{4^n-1}\epsilon_n\right)\rho^{[n]}\\
        &+ \left(\frac{1}{4^n-1}\epsilon_n\right)\sum_{\{\alpha_i\}}\left(\bigotimes_{i=1}^n\sigma_{\alpha_i}^i\right)\rho^{[n]}\left(\bigotimes_{i=1}^n\sigma_{\alpha_i}^i\right),
    \end{aligned}
\end{align}
where $\sigma_{\alpha_i}^i$ represents Pauli matrix $\sigma_{\alpha_i}$ applied on the $i$-th site.
The corresponding superoperator for this noise channel is written as
\begin{align}
    \begin{aligned}
        \mathcal{E}^{[n]} &= \left(1-\frac{4^n}{4^n-1}\epsilon_n\right)\bigotimes_{i=1}^n{\left[\sigma_{0}^i\otimes \sigma_{0}^{i*}\right]}\\
        &+ \left(\frac{1}{4^n-1}\epsilon_n\right)\bigotimes_{i=1}^n{\left[\sum_{\alpha_i = 0}^{3}\sigma_{\alpha_i}^i\otimes \sigma_{\alpha_i}^{i*}\right]},
    \end{aligned}
\end{align}
which can be represented by an MPO with $D = 2$, as shown in Fig. \ref{Fig: Dep_MPO}.

\section{Variational PEPO-inverse method}\label{App: PEPO}
\begin{figure*}
    \centering
    \subfigure[]{\includegraphics[width = 0.4\textwidth]{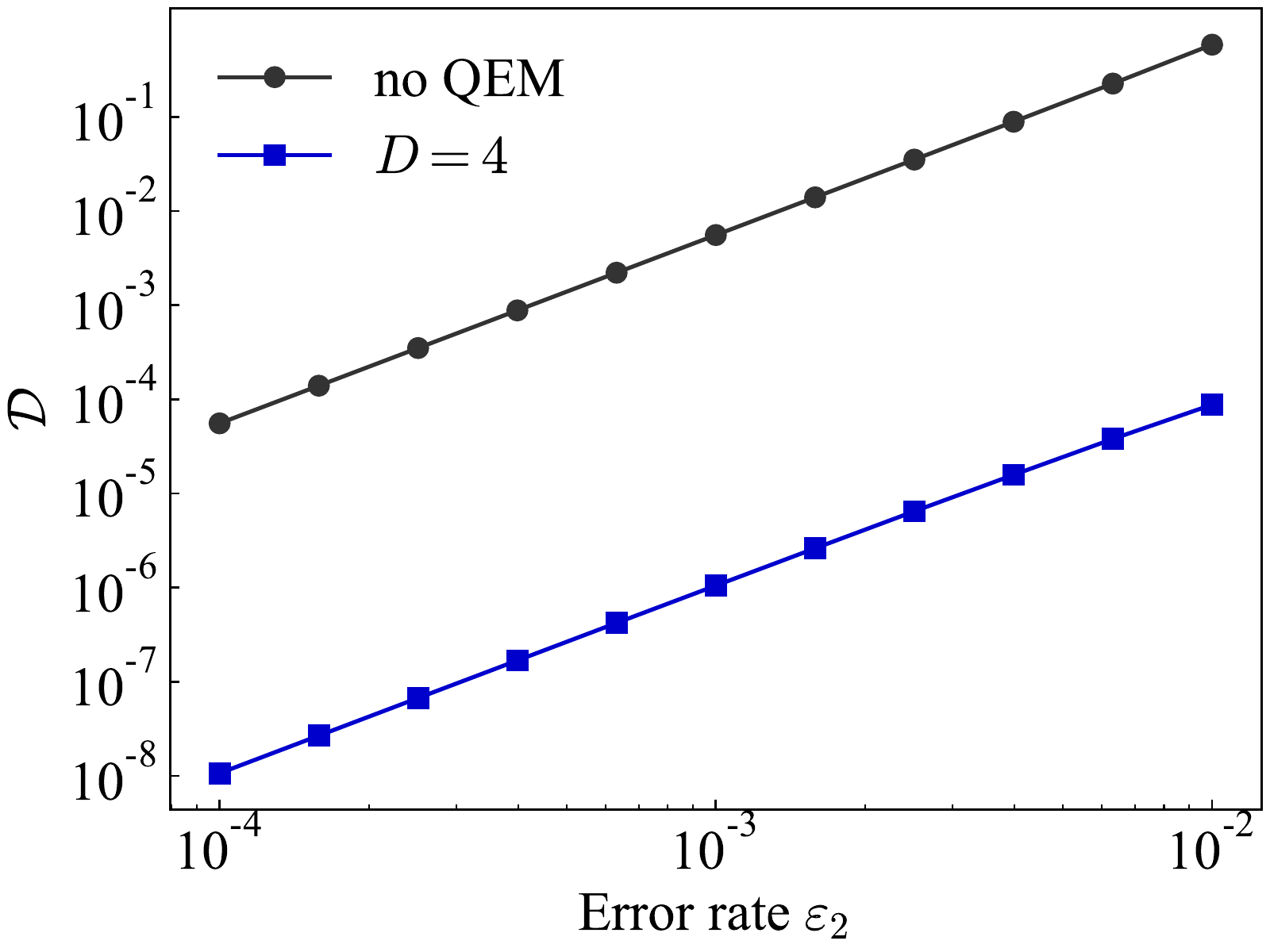}}
    \subfigure[]{\includegraphics[width = 0.4\linewidth]{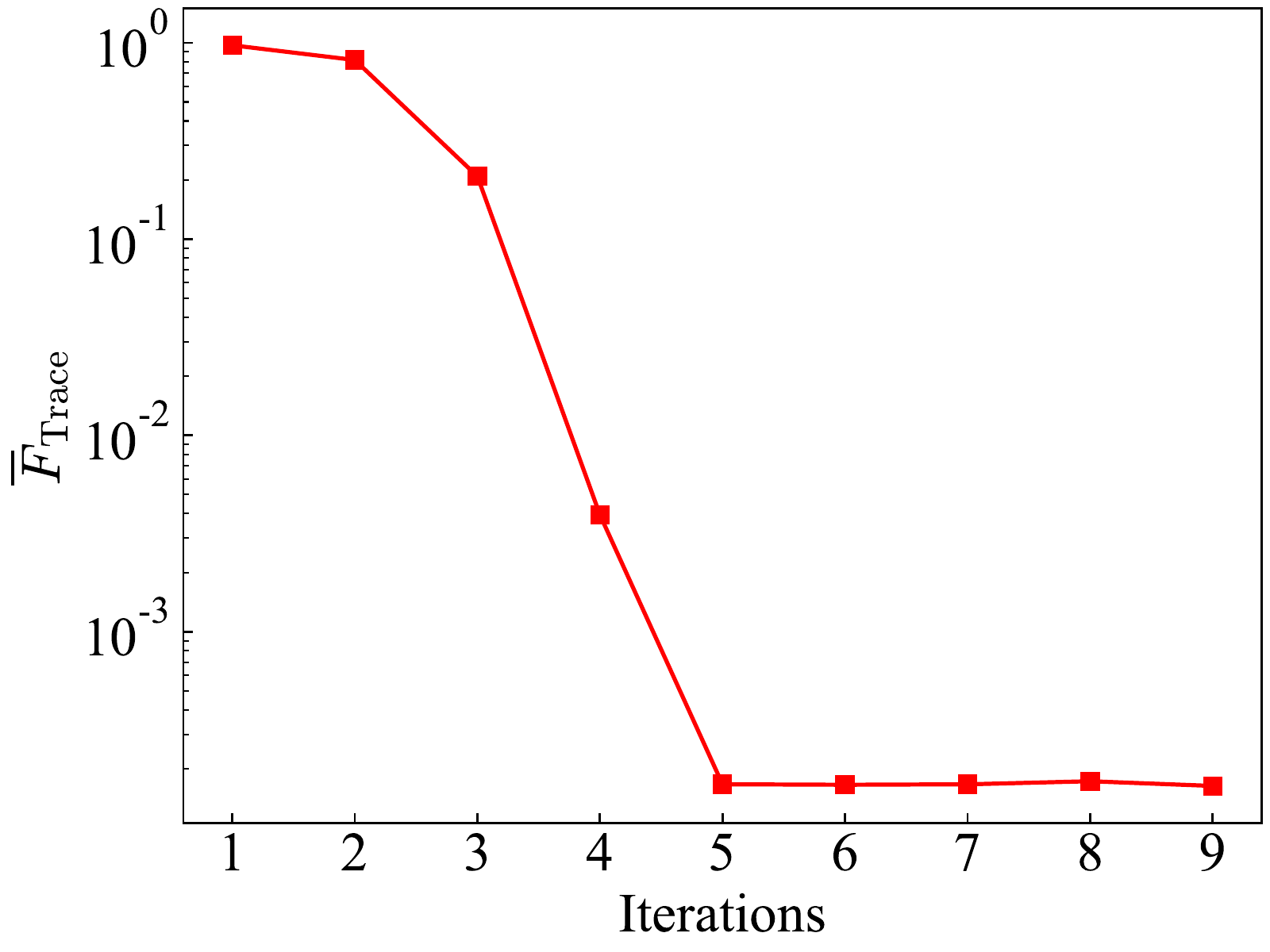}}
    \caption{(Color online) Variational PEPO-inverse method.
    (a) $\mathcal{D}(\mathcal{U}^{-1}_{\rm{PEPO}}\circ\mathcal{U}, \mathds{1})$ for different error rates.
    We benchmark $\mathcal{D}(\mathcal{U}, \mathcal{U}^{(0)})$ for the total noise effect in the original circuit, labeled as ``No QEM''.
    (b) $\overline{F}_{\text{Trace}}(\mathcal{U}^{\prime})$ during the iteration.
    }
    \label{Fig: Inv_PEPO}
\end{figure*}

We can directly generalize our MPO-inverse method to PEPO in 2D.
In this case, Eq. \eqref{Equ: Linear} remains unchanged, while the calculation of environment tensors involves contraction of 2D tensor networks, which can only be performed approximately.
Here we choose the standard contraction strategy for finite systems \cite{Orus2014}, i.e., we consider the 2D TN to be contracted as the evolution of a 1D MPS.
We contract local tensors and truncate the resulting MPS layer by layer, keeping its bond dimension no larger than $\chi$.
We set $\chi = D^2$ in our simulation.

We use a test circuit similar to the previous 1D case.
The even layer is the product of single-qubit gates randomly chosen from $\{Z, H, S, T\}$, while the odd layer is composed of CNOT gates.
These two-qubit gates are placed in different directions for adjacent odd layers.
Depolarizing noise is added after each gate with $\varepsilon_2 = 10 \varepsilon_1$, where $\varepsilon_i$ is the average error rate for $i$-qubit gates.
In Fig. \ref{Fig: Inv_PEPO} (a), we implement our PEPO-inverse method on a $6\times 6$, $\rm{depth} = 8$ noisy circuit for different error rates.
The trend is similar to the 1D case, indicating the validity of the PEPO-inverse method.

To verify the trace-preserving condition after taking the inverse, we also monitor $\overline{F}_{\text{Trace}}$ during the iteration process in Fig. \ref{Fig: Inv_PEPO} (b).
It is demonstrated that the trace-infidelity declines rapidly and converges to $\sim 10^{-4}$ in several iteration steps.
This result is intuitive since when $\mathcal{U}^{\prime} \mathcal{U}$ approaching $\mathds{1}$, $\mathcal{U}^{\prime}$ will approach the TP property of $\mathcal{U}^{-1}$.

\section{Finite-size effect}\label{App: Finite_size}
\begin{figure}
    \includegraphics[width=\linewidth]{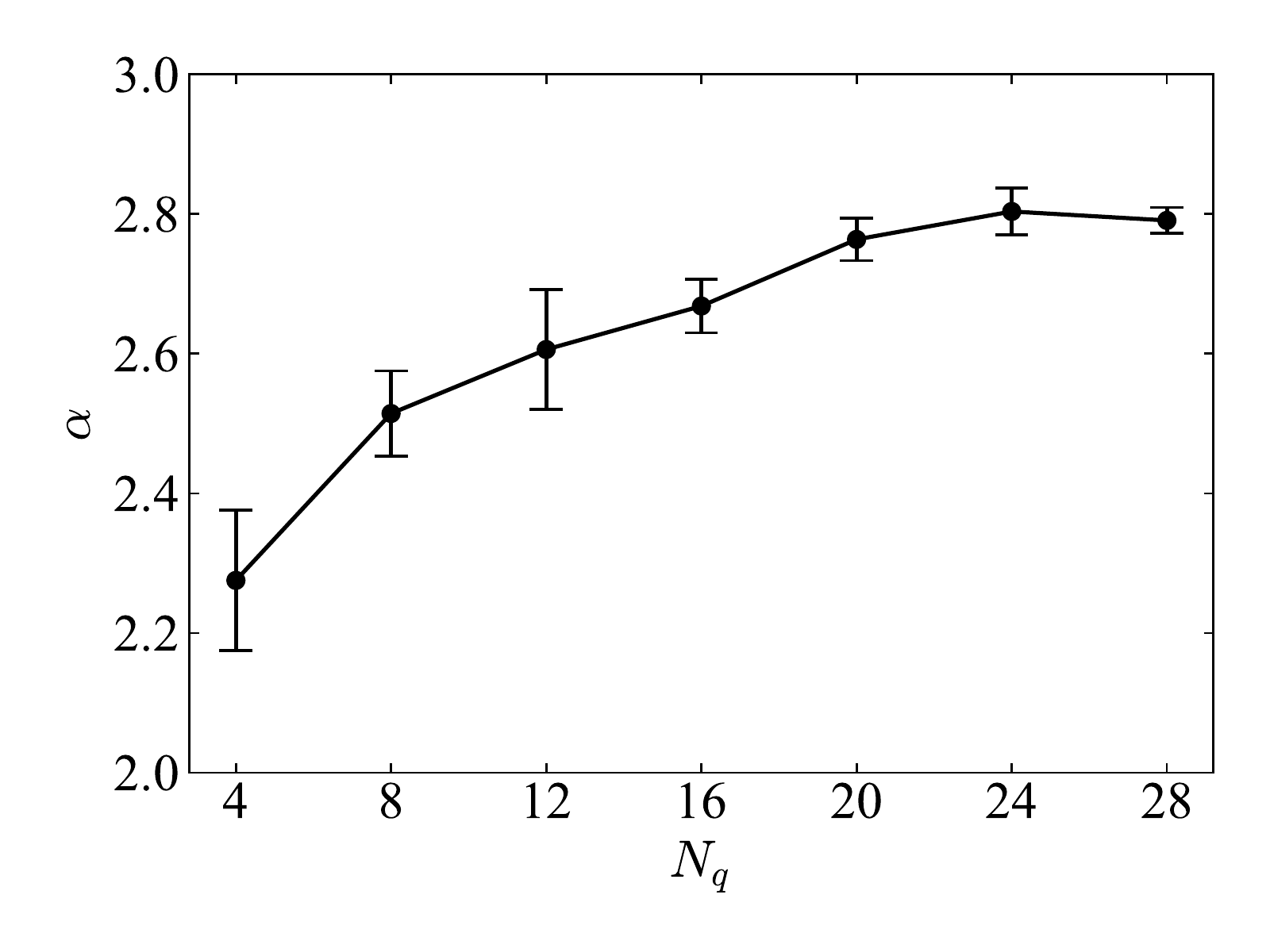}
    \caption{Exponents $\alpha$ in Eq. \eqref{equ: deep} for different $N_q$.
    The noise model is the same as that adopted in Sec. \ref{Sec: Deep}, where the average error rate for two-qubit gates is $\varepsilon_2 = 0.01$.}
    \label{Fig: Finite_size}
\end{figure}

In this section, we study the possible finite-size effect of our QEM method by directly fitting the exponent $\alpha$ defined in Eq. \eqref{equ: deep} for different $N_q$ and error-mitigated data.
It is shown in Fig. \ref{Fig: Finite_size} that $\alpha$ slightly increases for small $N_q$ and converges to $\alpha\approx 2.8$ at $N_q\sim 20$.
This demonstrates the validation of our discussions in Sec. \ref{Sec: Deep} where $N_q = 20$.
\bibliography{ref}
\end{document}